\documentclass[12pt]{article}
\pdfoutput=1
\usepackage{graphicx}
\usepackage{amsmath}
\allowdisplaybreaks
\usepackage{amssymb}
\usepackage{hyperref}

\usepackage{multirow}
\usepackage{cellspace}
\usepackage{caption}
\usepackage{mciteplus}
\graphicspath{{./}{./figures/}}

\newcommand{\MSbar}{\overline{\mbox{MS}}}
\newcommand{\msbar}{\overline{\mbox{\tiny{MS}}}}
\newcommand{\ri}{\mbox{i}}
\newcommand{\THDM}{2HDM}
\newcommand{\GeV}{\mbox{GeV}}
\newcommand{\rxi}{$\mathrm{R}_\xi$}
\newcommand{\re}{\mathrm{Re}}

\newcommand{\qgs}{{\tt{QGS}}}
\newcommand{\ew}{\mathrm{EW}}
\newcommand{\w}{\mathrm{weak}}
\newcommand{\kew}{K_{\ew}^{\mathrm{NLO}}}
\newcommand{\kbew}{\overline K_{\ew}^{\mathrm{NLO}}}
\newcommand{\kb}{\overline K}

\newcommand{\ds}{\displaystyle}

\newcommand{\Gf}{G_{F}}
\newcommand{\tb}{t_{\beta}}
\newcommand{\cab}{c_{\alpha\beta}}
\newcommand{\sab}{s_{\alpha\beta}}
\newcommand{\cw}{c_W}
\newcommand{\sw}{s_W}

\newcommand{\hl}{H_\mathrm{l}}
\newcommand{\hh}{H_\mathrm{h}}
\newcommand{\ha}{H_\mathrm{a}}
\newcommand{\hc}{H_\mathrm{c}}
\newcommand{\hpm}{H^{\pm}}
\newcommand{\gz}{G_0}

\newcommand{\hlb}{H_\mathrm{l,B}}
\newcommand{\hhb}{H_\mathrm{h,B}}
\newcommand{\hab}{H_\mathrm{a,B}}
\newcommand{\gzb}{G_\mathrm{0,B}}

\newcommand{\Mw}{M_W}
\newcommand{\Mz}{M_Z}
\newcommand{\mt}{m_t}

\newcommand{\Mhh}{M_{\hh}}
\newcommand{\Mhl}{M_{\hl}}
\newcommand{\Mha}{M_{\ha}}
\newcommand{\Mhc}{M_{\hc}}
\newcommand{\Msb}{M_\mathrm{sb}}

\newcommand{\gssp}{\Sigma_{SS'}}
\newcommand{\ghhhl}{\Sigma_{\hh\hl}}

\newcommand{\ggzha}{\Sigma_{\gz\ha}}

\newcommand{\gsspb}{\widetilde\Sigma_{SS'}}
\newcommand{\ghhhlb}{\widetilde\Sigma_{\hh\hl}}

\newcommand{\ggzhab}{\widetilde\Sigma_{\gz\ha}}

\newcommand{\thhhl}{t_{\hh\hl}}
\newcommand{\tgzha}{t_{\gz\ha}}

\newcommand{\da}{\delta\alpha}
\newcommand{\db}{\delta\beta}

\newcommand{\pOS}{OS}
\newcommand{\pst}{$p^*$}
\newcommand{\ch}{c^{\tiny\mathrm{\THDM}}_{\tiny H,q}}
\newcommand{\chl}{c^{\tiny\mathrm{\THDM}}_{\tiny\hl}}
\newcommand{\chh}{c^{\tiny\mathrm{\THDM}}_{\tiny\hh}}
\newcommand{\m}{\phantom{-}}

\begin{document}    

\begin{titlepage}
\noindent

\vspace{0.5cm}
\begin{center}
  \begin{Large}
\begin{center}
    \begin{bf}
      Electroweak corrections in the \THDM\\[-0.1cm]%
      for neutral scalar Higgs-boson\\[-0.1cm]%
      production through gluon fusion
    \end{bf}
    \end{center}
  \end{Large}
  \vspace{0.8cm}

    \begin{large}
      Laura Jenniches$\rm \, ^{a,\,}$%
      \footnote{\href{Laura.Jenniches@physik.uni-wuerzburg.de}{Laura.Jenniches@physik.uni-wuerzburg.de}},
      Christian Sturm$\rm \, ^{a,\,}$%
      \footnote{\href{Christian.Sturm@physik.uni-wuerzburg.de}{Christian.Sturm@physik.uni-wuerzburg.de}} 
      and 
      Sandro Uccirati$\rm \, ^{b,\,}$%
      \footnote{\href{uccirati@to.infn.it}{uccirati@to.infn.it}}
  \end{large}
  \vskip .7cm
        {\small {\em 
           $\rm ^a$ 
            Universit{\"a}t W{\"u}rzburg,\\
            Institut f{\"u}r Theoretische Physik und Astrophysik,\\
            Lehrstuhl f{\"u}r Theoretische Physik II,\\
            Campus Hubland Nord,\\
            Emil-Hilb-Weg 22,\\
            D-97074 W{\"u}rzburg, \\
            Germany}}\\[0.2cm]
        {\small {\em 
            $\rm ^b$ 
            Universit{\`a} di Torino e INFN,\\
            10125 Torino,\\
            Italy}}\\
        \vspace{0.8cm}
\vspace*{4cm}
{\bf Abstract}
\end{center}
\begin{quotation}
\noindent
We have computed the two-loop, electroweak corrections to the production
of a light and a heavy neutral, scalar Higgs-boson through the
important gluon fusion process in the Two-Higgs-Doublet Model.  We
provide our results in various renormalization schemes for different
scenarios and benchmark points, which will be valuable for experimental
studies at the LHC. We describe the technicalities of our two-loop calculation
and augment it by a phenomenological discussion. Our results are also
applicable to the gluonic neutral, scalar Higgs-boson decays.
\end{quotation}
\end{titlepage}
%
%
\section{Introduction\label{sec:Introduction}}
The Higgs boson, which has been discovered at the Large Hadron Collider
(LHC)~\cite{Aad:2012tfa,*Chatrchyan:2012xdj} with a mass of
$M_{h}=125.09\pm0.21\pm0.11$~GeV~\cite{Olive:2016xmw}, could be part of
an extended Higgs sector.  One of the simplest extensions of the
Standard Model (SM) is the Two-Higgs-Doublet Model~(\THDM), where an
extra scalar doublet is added.  Such an extension was introduced in
Ref.~\cite{Lee:1973iz} to provide an additional source of CP-violation,
which may contribute to explain the observed matter--anti-matter
asymmetry of the Universe. Special choices of the parameters, for
example in the Inert Model, can also provide a dark matter
candidate~\cite{Barbieri:2006dq,*Cao:2007rm}. 
Thus, the \THDM\ can assist to address problems which are not
solved within the SM. Phenomenological studies of the \THDM\ have been 
performed by the
ATLAS~\cite{%
 Aad:2015pla,
 Aaboud:2017rel,
*Aad:2015kna,
*Aad:2015uka,
*Aad:2015wra,
*Aad:2013dza,
*Potter:2013ura,
*ATLAS-CONF-2013-027%
} %
and
CMS~\cite{%
 CMS:2018lkl,
*Bondu:2016ait,
*Khachatryan:2016are,
*Khachatryan:2015baw,
*Khachatryan:2015tha,
*Khachatryan:2015qba,
*Khachatryan:2015lba,
*Khachatryan:2014jya
} %
collaborations.  
Several benchmark scenarios for the new parameters, which arise
due to the addition of a second scalar doublet, have been collected by
the LHC Higgs Cross Section Working Group in Ref.~\cite{deFlorian:2016spz}.

Extensions of the SM can strongly modify Higgs-boson production and
decay processes, which allows to perform exclusion studies for the new
parameters of such model extensions. This requires precise theoretical
predictions. For example, the addition of a sequential fourth generation
of heavy fermions increases the leading order~(LO) cross section of the
Higgs-boson production process through gluon fusion already by about a
factor of nine~\cite{Georgi:1977gs}. The next-to-leading order (NLO)
electroweak corrections in this model have been computed in
Refs.~\cite{Passarino:2011kv,Denner:2011vt}, which have helped to
exclude this model at the LHC.

In the CP-conserving \THDM, there are two
neutral, scalar Higgs bosons~$\hl$ and~$\hh$. One of them is considered to
be the SM-like Higgs boson, which has already been
discovered at the LHC. In addition, there are two charged Higgs
bosons~$H^{\pm}$ as well as a pseudoscalar Higgs boson~$\ha$.  A possible
set of new
free parameters of the extended Higgs sector are the masses of
the new Higgs bosons, $\Mhh$, $M_{H^{\pm}}$, $\Mha$ as well as the soft $Z_2$-breaking scale
$\Msb$ and two mixing angles $\alpha$ and $\beta$.
The mass of the light Higgs boson is here fixed to~$\Mhl\equiv M_{h}$.

Various important Higgs-boson production and decay processes have
already been studied at NLO in the \THDM.  For example, NLO electroweak
corrections to Higgs-boson production in Higgs strahlung and through vector-boson
fusion have been determined in the \THDM\ in
Refs.~\cite{Denner:2016etu,Denner:2017vms}. The NLO electroweak corrections to the decay
of the light Higgs boson of the \THDM\ into four fermions have been computed in
Ref.~\cite{Altenkamp:2017ldc,*Altenkamp:2017kxk}.
The dominant Higgs-boson production mechanism at the LHC proceeds via
gluon fusion. Its precise theoretical knowledge is thus a mandatory
task. In the SM, the complete NLO electroweak corrections for the most recent value
of the top-quark mass $m_t=173.1$~GeV~\cite{Olive:2016xmw} amount to
$5.1\%$~\cite{Actis:2008ug,Actis:2008ts}.
One can thus expect that the
electroweak corrections are also sizable in extensions of the SM.  In
this work, we will contribute to the effort of studying the \THDM\ by
computing the two-loop electroweak corrections to the
production of neutral, scalar Higgs bosons in gluon fusion.
On the one hand, it is important to know the NLO
electroweak corrections in the \THDM\ to Higgs-boson production in gluon
fusion for the already discovered Higgs boson.  First results in the
special case of the alignment limit~($\cos{(\alpha-\beta)}=0$)
have already been presented in Ref.~\cite{Denner:2016etu}. In addition to
the alignment limit,
we consider the more complicated, general case of
$\cos{(\alpha-\beta)}\ne0$ in this work.  On the other hand, it is important to
know the NLO electroweak corrections for the production of the heavy,
neutral, scalar Higgs boson $\hh$ of the \THDM, which we
present too. We also discuss the technicalities and obstacles which need to be
overcome in order to accomplish this calculation.
The results of our computations will be provided in different renormalization
schemes for the mixing angles $\alpha$ and $\beta$.

The structure of this paper is as follows: In Section~\ref{sec:model}, we
introduce the model in which we perform our
calculation. Section~\ref{sec:renormalization} describes the different
renormalization schemes which we use for the mixing angles. In
Section~\ref{sec:calculation}, we discuss the computational techniques 
which have been used in order to accomplish this aim; and in
Section~\ref{sec:results} we provide the numerical results. Finally, we
close with our summary and conclusion in
Section~\ref{sec:summary}. 
In the appendices, we provide supplementary information on the scale dependence
arising from the $\MSbar$ renormalization of the mixing angles $\alpha$ and 
$\beta$ as well as on the perturbative behaviour of the coupling constants of
the Higgs potential.

\section{The model\label{sec:model}}
We discuss the \THDM\ extension of the SM, which has two
complex, scalar doublet fields $\Phi_i$, $i=1,2$.
In the generic basis, they are parametrized through
\begin{equation}
  \label{eq:genericphi}
  \Phi_i=\begin{pmatrix}\phi_i^{+}\\{1\over\sqrt{2}}(v_i+\rho_i+\ri\eta_i)\end{pmatrix},
\end{equation}
where the $v_i$ are the vacuum expectation values (vev). 
We consider the \THDM\ with a discrete $Z_2$ symmetry under the transformation
$\Phi_1\to-\Phi_1$, $\Phi_2\to\Phi_2$ of the two Higgs doublets.  
This $Z_2$ symmetry is important, since it has also implications on the
Yukawa sector of the \THDM\, where it suppresses tree-level
flavour-changing neutral currents~(FCNC), which are experimentally
unobserved. Before coming to the Yukawa sector, let us complete the
discussion of the Higgs potential. The $Z_2$ symmetry requirement
also reduces the number of terms in the potential. 
However, we allow for a soft-breaking term~\cite{Kanemura:1999xf},
which does not induce the FCNC problem. 
With these constraints, the Higgs potential has the following form
\begin{eqnarray}
  V(\Phi_1,\Phi_2)&=&
   m_1^2\Phi_{1}^{\dagger}\Phi_{1}
  +m_2^2\Phi_2^{\dagger}\Phi_2
  -m_{12}^2\left(\Phi_{1}^{\dagger}\Phi_2+\Phi_2^{\dagger}\Phi_{1}\right)
  \nonumber\\
 &+&\frac{\lambda_1}{2}\left(\Phi_{1}^{\dagger}\Phi_{1}\right)^2
  +\frac{\lambda_2}{2}\left(\Phi_2^{\dagger}\Phi_2\right)^2
  +\lambda_3\left(\Phi_{1}^{\dagger}\Phi_{1}\right)\left(\Phi_2^{\dagger}\Phi_2\right)
  \nonumber\\
 &+&\lambda_4\left(\Phi_{1}^{\dagger}\Phi_2\right)\left(\Phi_2^{\dagger}\Phi_{1}\right)
  +\frac{\lambda_5}{2}\left[\left(\Phi_{1}^{\dagger}\Phi_2\right)^2
  +\left(\Phi_2^{\dagger}\Phi_{1}\right)^2\right].
\label{eq:hpgb}
\end{eqnarray}
The five couplings $\lambda_1$,...,$\lambda_5$ and the soft-breaking parameter
$m_{12}$ are taken to be real as well as the two masses $m_1$ and $m_2$.

For physical applications, we work in the physical basis,
where the sector of the potential that is 
quadratic in the scalar fields is diagonalized. This leads to the introduction
of the mass eigenstates through a change of basis 
\begin{equation}
  \begin{pmatrix}
    \rho_1\\\rho_2
  \end{pmatrix}
  =R(\alpha)
  \begin{pmatrix}
    H_h\\H_l
  \end{pmatrix},
  \quad
  \begin{pmatrix}
    \phi_1^{\pm}\\\phi_2^{\pm}
  \end{pmatrix}
  =R(\beta)
  \begin{pmatrix}
    G^{\pm}\\H^{\pm}
  \end{pmatrix},
  \quad
  \begin{pmatrix}
    \eta_1\\\eta_2
  \end{pmatrix}
  =R(\beta)
  \begin{pmatrix}
    \gz\\\ha
  \end{pmatrix}.
\end{equation}
Here, the symbols $\hl$, $\hh$, $\hpm\equiv \hc$ and $\ha$ are the fields of the
physical light, heavy, charged and  pseudoscalar Higgs bosons, which
receive the masses $\Mhl$, $\Mhh$, $\Mhc$ and $\Mha$, while
$\gz$ and $G^{\pm}$ are the neutral and charged would-be Goldstone bosons.

The rotation matrix that performs the diagonalization reads
\begin{equation}
  R(\gamma)=
  \begin{pmatrix}
    \cos{\gamma}&-\sin{\gamma}\\
    \sin{\gamma}&\phantom{+}\cos{\gamma}
  \end{pmatrix},\;
  \mbox{with}\,\,\gamma=\alpha\,\,\mbox{or}\,\,\beta.\label{eq:rotation}
\end{equation}
The five couplings in the potential can then be expressed in terms of
the Higgs-boson masses and the mixing angles. The explicit formulae are
given in Eqs.~(\ref{eq:lambda1})-(\ref{eq:lambda5}) of
Appendix~\ref{app:scale_plots}.

Finally, we also introduce the Higgs basis~\cite{Branco:1999fs} in which
only the neutral component of one of the two Higgs doublets, say the
first one, acquires a vev. This is achieved by the
linear combination
\begin{equation}
\begin{pmatrix}
  \Phi_{a}\\\Phi_{b}    
\end{pmatrix}
=R^{-1}(\beta)
\begin{pmatrix}
  \Phi_1\\\Phi_2
\end{pmatrix}.
\end{equation}
Expressing the fields $\Phi_{a,b}$ in terms of the fields in the physical
basis leads to
\begin{equation}
  \Phi_{a}=
  \begin{pmatrix}
    G^{+}\\
    {1\over\sqrt{2}}(v+\widehat{H}+\ri \gz)
  \end{pmatrix},
  \quad
  \Phi_{b}=
  \begin{pmatrix}
    H^{+}\\
    {1\over\sqrt{2}}(\widetilde{H}+\ri \ha)
  \end{pmatrix},
\end{equation}
where we defined the neutral Higgs-boson fields through the following linear combinations
\begin{equation}
\begin{pmatrix}
\widetilde{H}\\\widehat{H}    
\end{pmatrix}
=R^{-1}(\alpha-\beta)
\begin{pmatrix}
\hl\\\hh
\end{pmatrix}.
\end{equation}
It becomes apparent that only the neutral component of $\Phi_{a}$
acquires a vev, while in $\Phi_{b}$ it does not. We introduce the abbreviations
$\cab=\cos(\alpha-\beta)$ and $\sab=\sin(\alpha-\beta)$ as well as the
vev $v=(v_1^2+v_2^2)^{1/2}$. 
The choice of phases in the definition of the two doublets $\Phi_i$ in
Eq.~\eqref{eq:genericphi} leads to $\sab=-\sqrt{1-\cab^2}$. 
An additional property of the Higgs basis is that the Goldstone bosons are all
contained in one doublet, $\Phi_a$. 
Note, that the Higgs basis was particularly valuable in
Ref.~\cite{Denner:2016etu} in order to show that an $\MSbar$
renormalization of the mixing angle $\beta$ can become gauge dependent
when using popular tadpole renormalization schemes.
In the case of the alignment limit
\begin{equation}
  \label{eq:alignmentlimit}
  \cab\to0,\qquad\sab\to-1,
\end{equation}
the first doublet ($\Phi_{a}$) becomes SM-like and $\widehat{H}$ is
the light Higgs boson $\hl$ ($\widehat{H}\equiv\hl$) with SM-like
couplings to fermions and gauge bosons. All the other new physical
Higgs-boson fields ($H^{\pm}$, $\widetilde{H}\equiv-\hh$, $\ha$) are in the
second doublet $\Phi_{b}$. As a result, there is no mixing between
fields from one doublet with those of the other doublet in the alignment limit.
Vice versa, if
$\sab\to0$ (and $\cab\to1$)
the heavy Higgs boson $\hh$ has SM-like couplings.

In order to give masses to the fermions in the \THDM, one has different
options to construct Yukawa interactions between the fermionic and
scalar fields. In particular, one distinguishes four different types of
\THDM\ Yukawa terms. In type I, only $\Phi_2$ couples to fermions.
In type II, the down-type quarks as well as the
leptons couple to $\Phi_1$, while the up-type quarks couple to
$\Phi_2$. Then, in type Y (or type flipped), $\Phi_2$ couples to up-type quarks and charged
leptons, while $\Phi_1$ couples to down-type quarks, and finally, we have
type X (or type lepton specific), where $\Phi_2$ couples to quarks, while $\Phi_1$
couples to charged leptons. Within this work we focus on type II. This corresponds to the
configuration that is realized in the Minimal Supersymmetric SM.
As already mentioned in the beginning of this section, we impose a
discrete $Z_2$ symmetry in order to avoid tree-level
FCNC~\cite{Glashow:1976nt,*Paschos:1976ay}.  For the type II
\THDM, down-type quarks and charged leptons need to be odd under this
$Z_2$ transformation, i.e. $\Phi_1\to-\Phi_1$, $d_R\to-d_R$,
    $\ell_R\to-\ell_R$,
while all other fields remain unchanged. Here, the fields
$d_R$ and $\ell_R$ are the right-handed, down-type quarks and the 
right-handed charged leptons.  All fermions except for the top-quark are
taken to be massless such that our results are valid for all types of \THDM\
Yukawa terms.
This is only justified for small values of $\tb$.
Furthermore, we neglect flavour mixing in the
following.

As new input parameters of the extended Higgs sector, we 
have the masses of the heavy, charged and pseudoscalar Higgs bosons
\[
\Mhh,\quad
\Mhc,\quad
\Mha
\]
as well as the soft-breaking scale 
\begin{equation}
\Msb^2={m_{12}^{2}\over\cos{\beta}\sin{\beta}}
\end{equation}
and the mixing angles $\alpha$ and $\beta$, which we express through
\[
\cab\quad \mbox{and}\quad t_{\beta}=\tan\beta={v_2\over v_1}.
\]
Several limits of the new parameters can be defined.
Beside the already mentioned alignment limit, the decoupling limit will be studied
in this work. In this limit, not only $\cab$ is equal to zero, but in addition, all new mass scales of the \THDM\
are much larger than the electroweak scale~\cite{Gunion:2002zf,deFlorian:2016spz}.

\section{Renormalization of the mixing angles {\boldmath{$\alpha$}} and {\boldmath{$\beta$}}\label{sec:renormalization}}
In this section, we consider different renormalization
schemes for the mixing angles $\alpha$ and $\beta$. We distinguish
the $\MSbar$ renormalization, which leads to scale-dependent amplitudes,
and the on-shell, $p^*$ as well as two process-dependent schemes. The latter guarantee
scale-independent amplitudes. In the process considered in this work, the
parameter $\Msb$ does not enter at LO. Thus, it does not require renormalization
unless the running is taken into account in the $\MSbar$~renormalization
scheme.

In Section~\ref{sec:results}, we will show numerical results for these
renormalization schemes.
Having different renormalization schemes at hand can be
valuable in order to estimate the residual uncertainty due to
unknown higher order corrections.

\subsubsection*{$\MSbar$\ renormalization scheme}
An $\MSbar$\ renormalization of the mixing angles $\alpha$ and $\beta$
has been defined in Ref.~\cite{Denner:2016etu}. Here, the $\MSbar$\
counterterm $\delta\beta^{\msbar}$ is defined at the $\tau-H_a-\tau$
vertex, and the $\MSbar$\ counterterm $\delta\alpha^{\msbar}$ is defined
at the $\tau-H_l-\tau$ vertex.  The counterterms $\delta\alpha^{\msbar}$
and $\delta\beta^{\msbar}$ are required to render all amplitudes finite.
The treatment of tadpoles requires special care when using an
$\MSbar$\ renormalization of the mixing
angles $\alpha$ and $\beta$ in order to guarantee gauge-independent
amplitudes.  The tadpoles have been treated in the {\it{FJ}} tadpole
scheme, which has been introduced by Fleischer and Jegerlehner for the SM
in Ref.~\cite{Fleischer:1980ub} and which was extended for the \THDM,
and also for a general Higgs sector, in
Ref.~\cite{Krause:2016oke} and Ref.~\cite{Denner:2016etu}. 
In the SM, the FJ tadpole scheme corresponds to the $\beta_t$ scheme
of Ref.~\cite{Actis:2006ra}. In this approach, the
tadpole contributions are not absorbed into bare physical parameters,
which intrinsically assures gauge-independent physical counterterms.  A
renormalization of the soft-breaking scale $M_{sb}$ in the $\MSbar$\
scheme at the $H_c-H_h-H_c$ vertex was introduced in
Ref.~\cite{Denner:2016etu}. For an $\MSbar$\ renormalization
  of $M_{sb}$ see also Ref.~\cite{Krause:2016xku}.
At next-to-leading order, the counterterms
$\delta\alpha$ and $\delta\beta$ of the \THDM\ can be obtained from the off-diagonal
elements of the field-renormalization constants of the Higgs- and Goldstone-boson
fields~\cite{Kanemura:2015mxa}. In
particular, for a gauge-independent $\MSbar$ renormalization, this
relation reads
\begin{equation}
  \delta\alpha^{\msbar}={\delta Z^{\msbar}_{\hh\hl} -\delta Z^{\msbar}_{\hl\hh}\over4},\qquad
  \delta\beta^{\msbar} ={\delta Z^{\msbar}_{\gz\ha} -\delta Z^{\msbar}_{\ha\gz}\over4},
\end{equation}
where the $Z$-factors are defined by
\begin{equation}
  \left(
  \begin{matrix}
   \hhb \\ \hlb
  \end{matrix}
  \right)
  =
  \left(
  \begin{matrix}
    1+{1\over2}\delta Z_{\hh\hh}&{1\over2}\delta Z_{\hh\hl}\\
    {1\over2}\delta Z_{\hl\hh}&1+{1\over2}\delta Z_{\hl\hl}
  \end{matrix}
  \right)
  \left(
  \begin{matrix}
   \hh\\\hl
  \end{matrix}
  \right),
\end{equation}
\begin{equation}
  \left(
  \begin{matrix}
   \gzb \\ \hab
  \end{matrix}
  \right)
  =
  \left(
  \begin{matrix}
    1+{1\over2}\delta Z_{\gz\gz}&{1\over2}\delta Z_{\gz\ha}\\
    {1\over2}\delta Z_{\ha\gz}&1+{1\over2}\delta Z_{\ha\ha}
  \end{matrix}
  \right)
  \left(
  \begin{matrix}
   \gz\\\ha
  \end{matrix}
  \right),
\end{equation}
where the subscript $B$ denotes a bare
field. For the renormalization conditions of the $Z$-factors we refer to
Ref.~\cite{Denner:2016etu}.

The choice of an $\MSbar$ renormalization leads to scale-dependent
renormalized amplitudes. In order to account for the different scales in the new
Higgs sector, we use the average 
\begin{equation}
  \mu_0 = 
  \begin{cases}
    \ds\frac{\Mhl+\Mhh+\Mha+2\Mhc+\Msb}{6},&\mbox{if}\quad \Msb\neq 0\\[2ex]
    \ds\frac{\Mhl+\Mhh+\Mha+2\Mhc}{5},&\mbox{if}\quad \Msb= 0\\
  \end{cases}
  \label{eq:mu0}
\end{equation}
as a central renormalization scale, see Ref.~\cite{Altenkamp:2017ldc,*Altenkamp:2017kxk} for a
similar scale choice.

In addition to the explicit scale dependence of the amplitude, the renormalization
scale dependence, i.e.\ the running of the parameters, can be taken
into account by solving a system of coupled differential equations:

\begin{eqnarray}
  \frac{\partial\alpha}{\partial\ln\left(\mu^2\right)} & = &
  \mathrm{B}_{\alpha}\left(\alpha(\mu),\beta(\mu),\Msb(\mu)\right),\notag\\
  \label{eq:running}
  \frac{\partial\beta}{\partial\ln\left(\mu^2\right)} & = &
  \mathrm{B}_{\beta}\left(\alpha(\mu),\beta(\mu),\Msb(\mu)\right),\\
  \frac{\partial\Msb}{\partial\ln\left(\mu^2\right)} & = &
  \mathrm{B}_{\Msb}\left(\alpha(\mu),\beta(\mu),\Msb(\mu)\right).\notag
\end{eqnarray}
The explicit expression for the functions $\mathrm{B}_{\alpha}$, 
$\mathrm{B}_{\beta}$ and $\mathrm{B}_{\Msb}$ can directly be obtained from 
the pole part of the corresponding counterterms and are lengthy in
the non-alignment limit.

In the $\MSbar$ scheme the explicit values of the input parameter $\cab$
and $\tb$ need to be defined at a given scale $\mu$.  We use two
different choices of the scale at which we define these values: the
averaged scale $\mu_0$ of Eq.~(\ref{eq:mu0}) as well as the vacuum
expectation value $v$. For the latter choice we run the values from the
scale $v$
to the scale $\mu_0$ with the help of the renormalization group
equations (RGE)~(\ref{eq:running}). We perform the scale variation
$\mu_0/2$ and $2\mu_0$ and run the values of $\cab$ and $\tb$ to these
corresponding scales. All results in the $\MSbar$ scheme
in Section~\ref{sec:results} are presented with the
parameters defined at $\mu_0$. In Appendix~\ref{app:scaledep} we also compare the
different $\MSbar$ schemes for the $M^*$ benchmark scenarios,
which will be introduced in Section~\ref{sec:results}.

\subsubsection*{On-shell renormalization scheme}
We have also considered the on-shell~(OS) scheme,
where the renormalized mixing angles are defined
such that they diagonalize the radiatively corrected Higgs mass matrices
introduced in Section \ref{sec:model}. 
This connects the counterterms of the mixing angles
$\alpha$ and $\beta$ to the off-diagonal terms of the on-shell
two-point functions of the CP-even
Higgs bosons and of the CP-odd Higgs boson and
the neutral Goldstone boson, respectively.
The mixing two-point functions are evaluated on-shell, which guarantees
scale independence of the rotation matrices in Eq.~\eqref{eq:rotation},
see Ref.~\cite{Espinosa:2001xu,*Espinosa:2002cd}.
However, using finite parts of the mixing two-point functions to define the renormalization
constants of $\alpha$ and $\beta$ leads to gauge-dependent
renormalization constants in addition to the gauge-dependent tadpole terms discussed
in Ref.~\cite{Denner:2016etu}. The problem has been overcome in
Ref.~\cite{Krause:2016oke} using the pinch
technique~\cite{Espinosa:2001xu,*Espinosa:2002cd,Cornwall:1981zr,*Cornwall:1989gv,Papavassiliou:1989zd,*Degrassi:1992ue,*Degrassi:1993kn,*Papavassiliou:1994pr,*Papavassiliou:1995fq,*Papavassiliou:1995gs,*Papavassiliou:1996zn,*Kniehl:2000kk,*Binosi:2002ez,*Binosi:2002bs}. Recently, it has been treated
in a more general framework in Ref.~\cite{Denner:2017vms} within the background field method (BFM),
leading to the same result. In the pinch technique the two-point
    function $\gsspb\left(p^2\right)$ of two scalars $S$ and $S'$ for
the \rxi-gauge with $\xi=1$ is given by 
\begin{equation}
\gsspb\left(p^2\right)=\gssp^\mathrm{1PI}\left(p^2\right)+\gssp^\mathrm{add}\left(p^2\right),
\label{eq:SE_pt}
\end{equation}
where $\gssp^\mathrm{1PI}\left(p^2\right)$ is the one-particle 
irreducible (1PI) mixing two-point function. In the \THDM\ the additional contribution
    $\gssp^\mathrm{add}\left(p^2\right)$ reads~\cite{Krause:2016oke}
\begin{eqnarray}
\ghhhl^\mathrm{add}\left(p^2\right) &=&
\frac{g^2\sab\cab}{32\pi^2\cw^2}
\left(p^2-\frac{\Mhh^2+\Mhl^2}{2}\right)
\Big\{
B_0\left(p^2;\Mz^2,\Mz^2\right)-B_0\left(p^2;\Mz^2,\Mha^2\right)\notag\\
&&\qquad+2\cw^2\left[B_0\left(p^2;\Mw^2,\Mw^2\right)-B_0\left(p^2;\Mw^2,\Mhc^2\right)\right]
\Big\},
\label{eq:pt_add}\\
\ggzha^\mathrm{add}\left(p^2\right) &=&
\frac{g^2\sab\cab}{32\pi^2\cw^2}
\left(p^2-\frac{\Mha^2}{2}\right)
\Big\{
B_0\left(p^2;\Mz^2,\Mhl^2\right)-B_0\left(p^2;\Mz^2,\Mhh^2\right)
\Big\},\notag
\end{eqnarray}
with the cosine of the weak mixing
angle~$\cw$, the gauge coupling $g$ and the one-loop, scalar, two-point
integral $B_0$~\cite{tHooft:1978jhc,*Passarino:1978jh}.

The counterterms to the mixing angles are then given by
\begin{equation}
\begin{aligned}
\da &=&
\frac{
\re\left[\ghhhlb\left(\Mhh^2\right)+\ghhhlb\left(\Mhl^2\right)+2\thhhl\right]
}{
2\left(\Mhh^2-\Mhl^2\right)
},\\
\db &=&\!\!\!-\;\;
\frac{
\re\left[\ggzhab\left(0\right)+\ggzhab\left(\Mha^2\right)+2\tgzha\right]
}{
2\Mha^2
},
\end{aligned}
\end{equation}
where the tadpole counterterms $\thhhl$ and $\tgzha$ are given in the Appendix of Ref.~\cite{Denner:2016etu}.
In an alternative on-shell scheme, the counterterm $\db$ can also be
defined through the charged Higgs-boson--Goldstone-boson mixing two-point
functions. 

These schemes can be expected to produce small perturbative corrections to physical
observables, because the finite
counterterms of the mixing angles $\alpha$ and $\beta$ cancel
large terms originating from the Z-factors of the neutral, scalar sector and
in the pseudoscalar/charged sector as mentioned in Ref.~\cite{Denner:2017vms}.
\subsubsection*{$\boldsymbol{p^*}$ renormalization scheme}
The $p^*$ scheme is derived from the same mixing two-point functions of
Eq.~\eqref{eq:SE_pt}, but computed at the external momentum
\begin{equation}
(p^*)^2=\frac{M_S^2+M_{S'}^2}{2},
\end{equation}
such that the rotation matrices in Eq.~\eqref{eq:rotation} also remain 
scale-independent, see Ref.~\cite{Espinosa:2001xu,*Espinosa:2002cd}. 
The additional terms in Eqs.~\eqref{eq:pt_add} vanish and the
two-point functions in Eq. \eqref{eq:SE_pt} are equal to the
1PI two-point functions in the \rxi-gauge for $\xi=1$. This results in the
counterterms of the mixing angles
\begin{equation}
\begin{aligned}
\da &=
\frac{\displaystyle
\re\left[\ghhhlb\left(\frac{\Mhh^2+\Mhl^2}2\right)+\thhhl\right]
}{
\Mhh^2-\Mhl^2
}
=
\frac{\displaystyle
\re\left[\ghhhl\left(\frac{\Mhh^2+\Mhl^2}2\right)+\thhhl\right]_{\xi=1}
}{
\Mhh^2-\Mhl^2
},\\
\db &=
-\frac{\displaystyle
\re\left[\ggzhab\left(\frac{\Mha^2}2\right)+\tgzha\right]
}{
\Mha^2
}
=
-\frac{\displaystyle
\re\left[\ggzha\left(\frac{\Mha^2}2\right)+\tgzha\right]_{\xi=1}
}{
\Mha^2
}.
\end{aligned}
\end{equation}

\subsubsection*{Process-dependent renormalization scheme 1}
\begin{figure}
\begin{center}
\begin{minipage}{2.5cm}
\includegraphics[width=2.5cm]{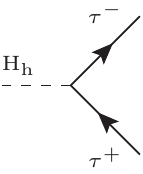}
\end{minipage}
\begin{minipage}[b]{4cm}
\[\leftrightarrow\quad {\ri\,e\,m_{\tau}\over2\,\sw\,\Mw}(\tb\,\sab\,-\,\cab)\]
\end{minipage}
\hspace*{2cm}
\begin{minipage}{2.5cm}
\includegraphics[width=2.5cm]{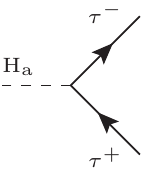}
\end{minipage}
\begin{minipage}[b]{4cm}
\[
\leftrightarrow\quad {e\,m_{\tau}\over2\,\sw\,\Mw}\,\gamma_5\,\tb
\]
\end{minipage}\\
\begin{minipage}{2.5cm}
\includegraphics[width=2.5cm]{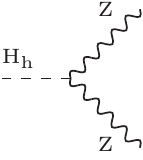}
\end{minipage}
\begin{minipage}[b]{4cm}
\[
\leftrightarrow\quad{\ri\, e\over\sw\,\cw}\,{\Mw\over\cw}\,\cab\,g^{\mu\nu}
\]
\end{minipage}
\end{center}
  \caption{The vertices that are used for the determination of the
    counterterms $\delta\beta$, $\delta\alpha$ or $\delta\phi$ with $\phi=\alpha-\beta$ are shown with their
    corresponding Feynman rules. The symbol $e$ is the elementary
    electric charge, $\Mw$ ($m_\tau$) is the mass of the $W$ boson
    ($\tau$ lepton) and $\sw=\sqrt{1-\cw^2}$.\label{fig:vertices}}
\end{figure}

A process-dependent renormalization of the parameters $\alpha$ and
$\beta$ can be obtained by imposing renormalization conditions on the physical decay
processes $H_h\to\tau^+\tau^-$ and $H_a\to\tau^+\tau^-$~\cite{Krause:2016oke}.
The two counterterms $\delta\alpha$ and $\delta\beta$ are fixed by the
requirement that their NLO corrected partial decay width
$\Gamma^{\mbox{\tiny{NLO}}}_{\w}$, which contains
{\it{weak}} corrections only, is equal to the LO partial decay
width~$\Gamma^{\mbox{\tiny{LO}}}$, i.e.
\[
  \Gamma^{\mbox{\tiny{NLO}}}_{\w}(H_h\to\tau^+\tau^-)=\Gamma^{\mbox{\tiny{LO}}}(H_h\to\tau^+\tau^-)
  \quad\mbox{and}\quad
  \Gamma^{\mbox{\tiny{NLO}}}_{\w}(H_a\to\tau^+\tau^-)=\Gamma^{\mbox{\tiny{LO}}}(H_a\to\tau^+\tau^-).
\]
Since both vertices ($\tau-H_h-\tau$ and $\tau-H_a-\tau$) depend on 
$\tb$, see Fig.~\ref{fig:vertices} for
illustration, the determination of both counterterms
$\delta\alpha$ and $\delta\beta$ requires the solution of a linear
system of equations. First, $\delta\beta$ can be determined from
the $\tau-H_a-\tau$ vertex, which depends only on $\tb$. Then, the result needs to
be inserted into the $\tau-H_h-\tau$ vertex for the determination of
$\delta\alpha$.
We refer to this scheme as proc1 in the following.

If the additional \THDM\ particles are discovered, one can expect that these
two processes can be measured precisely due to the leptonic final
states. Therefore, they are in principle well suited for the
determination of the parameters $\alpha$ and $\beta$. Due to the choice
of the renormalization condition, no higher order electroweak corrections
are present in the determination of $\alpha$ and $\beta$.
Process-dependent renormalization may lead to
large perturbative corrections, since the higher order
corrections to the processes used for renormalization will appear in all other
processes. Such a behaviour was, for example, observed for the decay
process $\hpm\to W^{\pm}\hl$ in Ref.~\cite{Krause:2016oke}. For
light and heavy 
Higgs-boson production through gluon fusion, however, process-dependent renormalization
leads to results similar to
those obtained in the \pOS\ or \pst\ scheme, as we will see in
Section~\ref{sec:results}.

\subsubsection*{Process-dependent renormalization scheme 2}
Unlike the previously discussed physical renormalization conditions, which 
rely on the partial decay widths of two leptonic Higgs-boson 
decays, we consider a process dependent renormalization, which is
based on the two vertices $\tau-H_a-\tau$ and $Z-H_h-Z$ in the following.  Studying a
potential decay of a heavy Higgs boson into two $Z$-bosons is
experimentally less clean, since the unstable $Z$-bosons can further
decay into a variety of other particles. In addition, its LO contribution
is very small in physically relevant scenarios close to the alignment
limit. Nevertheless, it turns out, as we
will see in Section~\ref{sec:results}, that in the calculation of the processes
$g+g\to\hl$ and $g+g\to\hh$,
this scheme leads to higher order corrections of moderate size.

As renormalization condition, we require that the purely
weak corrected $\tau-H_a-\tau$ vertex is equal to its leading order
value in order to fix the counterterm $\delta\beta$. We impose a similar
condition on the $Z-H_h-Z$ vertex in order to fix the counterterm
$\delta\phi$.  Note that due to our choice of parameters,
$\tb$ and $\cab$, it is more convenient to renormalize
$\phi=\alpha-\beta$ rather than $\alpha$. 
The $Z-H_h-Z$ vertex is proportional to $g^{\mu\nu}$ at LO, see
Fig.~\ref{fig:vertices}. Computing higher order corrections to this 
vertex 
leads to a richer tensor structure, which also contains
combinations of the four-momenta
of the external $Z$-bosons. As renormalization condition for
the counterterm $\delta\phi$, we require that the coefficient in front
of $g^{\mu\nu}$ of the $Z-H_h-Z$ amplitude is equal to its leading order
value similar to the renormalization of the electric charge in QED.
The coefficient in front of $g^{\mu\nu}$ can be extracted from
the amplitude by using a suitable projector.  This scheme will be
called proc2 in the following.  The use of the vertices
$\tau-H_a-\tau$ and $Z-H_h-Z$ has the advantage that one has a separate
condition for each of the two counterterms $\delta\beta$ and
$\delta\phi$, i.e.\ the value of $\delta\phi$ is not influenced by the
renormalization condition for $\delta\beta$, contrary to the
previously discussed proc1 scheme.

\section{Calculation\label{sec:calculation}}
The partonic, leading order cross section for the Higgs-boson production
process through gluon fusion reads
\begin{eqnarray}
\nonumber
\sigma_{\mbox{\tiny{\THDM}}}^{\mbox{\tiny{LO}}}(g+g\to H)&=&
{G_F M_H^2\alpha_s^2\over128\sqrt{2}\pi}\left|A_H^{\mbox{\tiny{LO}}}\right|^2\delta(s-M_H^2)\\
&\equiv&\hat{\sigma}^{\mbox{\tiny{LO}}}M_H^2\delta(s-M_H^2),
\label{eq:ggHLOSigma}
\end{eqnarray}
where $s$ is the square of the sum of the external gluon~($g$) momenta,
$\alpha_s$ is the strong coupling constant 
and $G_F$ is the Fermi coupling constant. The leading order amplitude
$A^{\mbox{\tiny{LO}}}$ reads
\begin{equation}
\label{eq:AHggLO}
A_H^{\mbox{\tiny{LO}}}=\sum_q\ch{1\over\tau_q}\left[1+\left(1-{1\over\tau_q}\right)f(\tau_q)\right]
\quad\mbox{with}\quad \tau_q=\frac{M_H^2}{4\*M_q^2},
\end{equation}
where the index $q$ runs over all quark flavours and 
\begin{equation}
\label{eq:ftau}
f(\tau_q)=
\begin{cases}
\arcsin^2\!\sqrt{\tau_q},& \mbox{if } \tau_q\leq 1,\mbox{ i.e. }q=t\\
-{1\over4}\left[\ln\left({1+\sqrt{1-1/\tau_q}\over1-\sqrt{1-1/\tau_q}}\right)-\ri\pi\right]^2,
& \mbox{if } \tau_q> 1,\mbox{ i.e. }q=u,d,s,c,b.
\end{cases}
\end{equation}
The symbol $H$ in Eq.~(\ref{eq:ggHLOSigma}) stands for the external
Higgs boson, which can be either the light ($H=\hl$) or the heavy
($H=\hh$) neutral Higgs boson of the \THDM; equally, $M_H$ is the
Higgs-boson mass, which can be either $M_H=\Mhl$ or
$M_H=\Mhh$. The symbol $M_q$
is the mass of the internal quark, which in general can be either the
up~($u$), down~($d$), strange~($s$), charm~($c$), bottom~($b$) or
top~($t$) quark.
The coefficient $\ch$ in Eq.~\eqref{eq:AHggLO} originates
from the Higgs-boson--quark coupling and is in the \THDM, type II 
different for up- and down-type quarks and also different for the production of a light or a heavy, neutral
Higgs boson: 
\begin{eqnarray}
\label{eq:chl}
&&c^{\tiny\mathrm{\THDM}}_{\tiny\hl,q}=\phantom{+}{\cab}/{\tb} - \sab
\;\;\;\;
\mbox{and}
\;\;\;\;
c^{\tiny\mathrm{\THDM}}_{\tiny\hh,q}=\cab + {\sab}/{\tb}
\;\;\;\;\mbox{for}\;\; q\in\{u,c,t\},\\\nonumber
&&c^{\tiny\mathrm{\THDM}}_{\tiny\hl,q}=-{\cab}\;{\tb} - \sab
\;\;\;\;\,
\mbox{and}
\;\;\;\;
c^{\tiny\mathrm{\THDM}}_{\tiny\hh,q}=\cab - {\sab}\;{\tb}
\;\;\;\;\,\mbox{for}\;\; q\in\{d,s,b\}.
\end{eqnarray}
The production of the SM Higgs boson corresponds to the case where
$\ch$ is equal to one. 
We consider only the top quark as massive
and all other fermions as massless, i.e. only the case $q=t$ contributes
to Eq.~(\ref{eq:ggHLOSigma}).
In this case only $c^{\tiny\mathrm{\THDM}}_{\tiny\hl,t}$ or
$c^{\tiny\mathrm{\THDM}}_{\tiny\hh,t}$ of Eq.~(\ref{eq:chl}) contributes
    to Eq.~(\ref{eq:AHggLO}). For simplicity we will drop the label $t$
in the following,
i.e. $c^{\tiny\mathrm{\THDM}}_{\tiny\hl}\equiv c^{\tiny\mathrm{\THDM}}_{\tiny\hl,t}$
and  
$c^{\tiny\mathrm{\THDM}}_{\tiny\hh}\equiv c^{\tiny\mathrm{\THDM}}_{\tiny\hh,t}$. 
 In particular,
$c^{\mbox{\tiny{\THDM}}}_{\hl}$ of Eq.~(\ref{eq:chl}) becomes one,
i.e. SM-like, in the alignment limit given in Eq.~(\ref{eq:alignmentlimit}).
One can also find a combination of $\cab$ and
$\tb$ such that $c^{\tiny\mathrm{\THDM}}_{\tiny H}$ becomes arbitrary small or
even zero. In this case, the two-loop, electroweak corrected
process becomes the true leading order process.

There are no real corrections
which have to be determined when computing the NLO electroweak
corrections for this process. Consequently, one can straightforwardly apply the
results for the electroweak percentage correction, which has been obtained for the
cross section of the production process $g+g\to H$ in Eq.~(\ref{eq:ggHLOSigma}),
to the partial decay
width of the process $H\to g+g$. The partial decay width and the production cross section are related by
\begin{equation}
\Gamma_{\mbox{\tiny{LO}}}(H\to g+g)={8M_H^3\over\pi^2}\hat{\sigma}_{\mbox{\tiny{LO}}}(g+g\to H).
\end{equation}
The Feynman rules for the \THDM\ of type II, which has been defined in
Section~\ref{sec:model}, have been produced in an automated way with the help of
{\tt{FeynRules}}~\cite{Alloul:2013bka}. The one- and two-loop diagrams have been
calculated with \qgs, which is an extension of {\tt{GraphShot}}~\cite{Actis}.
It generates the Feynman diagrams with {\tt{QGRAF}}~\cite{Nogueira:1991ex}
and performs algebraic manipulations of the amplitudes and loop integrals
with the help of {\tt{Form}}~\cite{Vermaseren:2000nd,*Kuipers:2012rf}. The program
\qgs\ performs the standard Dirac algebra operations and projects the expressions onto
form factors. In addition, it removes reducible scalar products and uses
integration-by-parts identities in order to simplify tadpole contributions.
Finally, it reduces the number of loop integrals
by means of symmetrization. Further details regarding these techniques can
be found in Ref.~\cite{Actis:2008ts}.
After these manipulations, one remains with a two-loop amplitude that 
requires numerical evaluation, which is done with a Fortran code. In
particular, the two-loop integrals are evaluated in Feynman-parametric
space with the help of a Fortran library.  For the numerical integral
evaluation of the two-loop massive diagrams, the library uses the methods
of Ref.~\cite{Passarino:2001wv,*Passarino:2001jd} for self-energies and
of
Refs.~\cite{Actis:2008ts,Ferroglia:2003yj,*Passarino:2006gv,Actis:2004bp}
for vertex functions.

\begin{figure}[!h]
\begin{center}
\includegraphics[width=4.5cm]{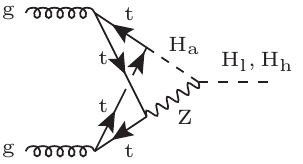}
\hspace*{0.3cm}
\includegraphics[width=4.5cm]{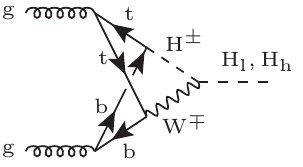}
\hspace*{0.3cm}
\includegraphics[width=4.5cm]{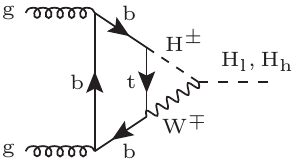}
\vspace{0.3cm}
\end{center}
  \caption{Sample diagrams that do not appear in light Higgs-boson production in the alignment limit are shown.
    In the second and third diagram, the bottom quarks are 
    considered to be massless. For light Higgs-boson production the
    contribution to the amplitude originating from these three diagrams
    is proportional to $\cos{(\alpha-\beta)}$. This is the reason why
    they vanish in the alignment limit.
\label{fig:vh}}
\end{figure}

Compared to the production of a light, neutral, SM-like
Higgs boson $\hl$ in gluon fusion in the alignment limit, which has been addressed
already briefly in Ref.~\cite{Denner:2016etu}, the non-alignment limit
case and the production of a heavy, neutral Higgs boson $\hh$ is
computationally more involved.
While the calculation of the integrals for the production
of the SM-like Higgs boson $\hl$ in the alignment limit can be completely
traced back to the calculation of the pure SM case, which was discussed
in Refs.~\cite{Actis:2008ug,Actis:2008ts}, new integral structures appear in the
general case. 
They are generated by new diagrams, a sample of which is shown in Fig.~\ref{fig:vh}. 
The first diagram  
is non-planar and leads to a new rank 2
integral. In the SM and in the alignment limit, this diagram only exists with
two Z-bosons. 
The same type of diagram exists for the W-bosons and the charged Higgs bosons (second diagram of Fig.~\ref{fig:vh}). Since different types of fermions appear
in this diagram, it leads to rank 3 integrals, and hence, the calculation
becomes even more involved. In the SM, it was possible to cancel the rank 3 integrals
since there were fewer different bosonic masses in the denominators.
A new planar diagram, which appears in the general case of the
\THDM, is depicted at the rightmost side of Fig.~\ref{fig:vh}.
After performing the obvious reduction as explained in
Ref.~\cite{Actis:2008ts}, some diagrams lead to collinear
    divergent structures; the new ones, not present in the alignment
    limit, are shown in Fig.~\ref{fig:Vcoll}. 
In particular, the second non-planar
    diagram of Fig.~\ref{fig:vh} leads to the more
complicated collinear structure depicted in the second diagram of Fig.~\ref{fig:Vcoll}.
More details regarding the cancellation of these new collinear singularities
are given in Section~\ref{sec:tests}.

In Section~\ref{sec:results}, we provide numerical results for 
the NLO, electroweak corrections to the production processes $g+g\to
\hl$ and $g+g\to \hh$ for various \THDM\ scenarios. We define the NLO corrections in terms
of the K-factor
\begin{eqnarray}
  |A_H^{\mbox{\tiny NLO,EW}}|^2 &=& |A_H^{(1)}+A_H^{(2)}|^2\notag\\&=&|A_H^{(1)}|^2+2\re\left(A_H^{(1)}A_H^{(2)*}\right)+\mathcal{O}\left(\left(\Gf\Mw^2\right)^2\right)\equiv|A_H^{(1)}|^2\kew,
\label{eq:Kfactor}
\end{eqnarray}
using the one-loop $A_H^{(1)}$ and two-loop $A_H^{(2)}$ contributions to the
amplitude with the normalization defined in Eq.~\eqref{eq:ggHLOSigma}.

If the coefficient $\chl$ or $\chh$ of Eq.~\eqref{eq:chl} becomes very small,
the one-loop amplitude $A_H^{(1)}$ is very small too, and the 
two-loop amplitude $A_H^{(2)}$ becomes the true leading order result.
As discussed in the context of Higgs-boson production and decay
with a fourth fermion generation~\cite{Denner:2011vt}, neglecting the term $|A_H^{(2)}|^2$ is no longer justified in this case, and we define a new K-factor
\begin{equation}
  |A_H^{\mbox{\tiny NLO,EW}}|^2 = |A_H^{(1)}+A_H^{(2)}|^2=|A_H^{(1)}|^2+2\re\left(A_H^{(1)}A_H^{(2)*}\right)+|A_H^{(2)}|^2\equiv|A_H^{(1)}|^2\kbew.
\label{eq:Kbfactor}
\end{equation}
Whenever $\kbew$ differs significantly from $\kew$, $\kbew$ will be used in Section~\ref{sec:results}.

\subsection{Analytical and numerical tests\label{sec:tests}}
We have performed several analytical as well as numerical tests to validate the
new components of \qgs, which are the \THDM\ Feynman rules, automatic generation
of Feynman diagrams and the appearances of new rank 2 and 3 integrals.

The ultraviolet (UV) structure of the new integrals as well as the consistency
of the \THDM\ Feynman rules has been tested by extracting all UV poles
in dimensional regularization,
as explained in Ref.~\cite{Actis:2008ts}, and verifying their cancellation
analytically.

\begin{figure}[!h]
\begin{center}
\includegraphics[width=4cm]{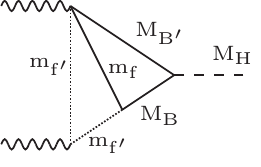}
\hspace*{2.3cm}
\includegraphics[width=4cm]{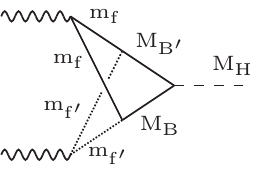}
\end{center}
  \caption{Collinear structures that do not appear in the alignment
    limit are shown. Massless particles are marked as wavy or dotted lines,
    while solid lines denote massive particles.
\label{fig:Vcoll}}
\end{figure}
Furthermore, the cancellation of collinear singularities can also be 
used to validate the implementation of the Feynman rules and the 
calculation of the new integral structures.
Collinear singularities arise in some diagrams, if the external 
gluons couple to light fermions, but they have to cancel when summing up 
the contributions from all diagrams. 
These singularities have been regularized by fictitious small fermion 
masses and become manifest in terms of linear and quadratic logarithms. 
The two-loop electroweak amplitude is subdivided into the contribution 
that comes from the first and second generation of fermions and the
contribution that originates from the third generation.  The
cancellation of the collinear divergent logarithms, which arise from the
first and second generation of fermions, has been verified analytically.
For the third generation of fermions, two new collinear structures, 
originating from the second and third diagram in Fig.~\ref{fig:vh}, 
are presented in Fig.~\ref{fig:Vcoll}.
Their analytic cancellation requires new integration-by-parts identities and
a more complicated partial-fraction decomposition due to new denominators.
All quadratically collinear divergent logarithms can then be cancelled analytically.
The linearly collinear divergent logarithms have been treated in a
numerical approach. Here, the logarithm is extracted analytically,
but its coefficient is evaluated numerically. We have 
verified that the sum of all these coefficients cancels 
numerically.

\begin{figure}[!h]
  \begin{center}
    \includegraphics[width=0.6\textwidth,bb=0 0 614 417]{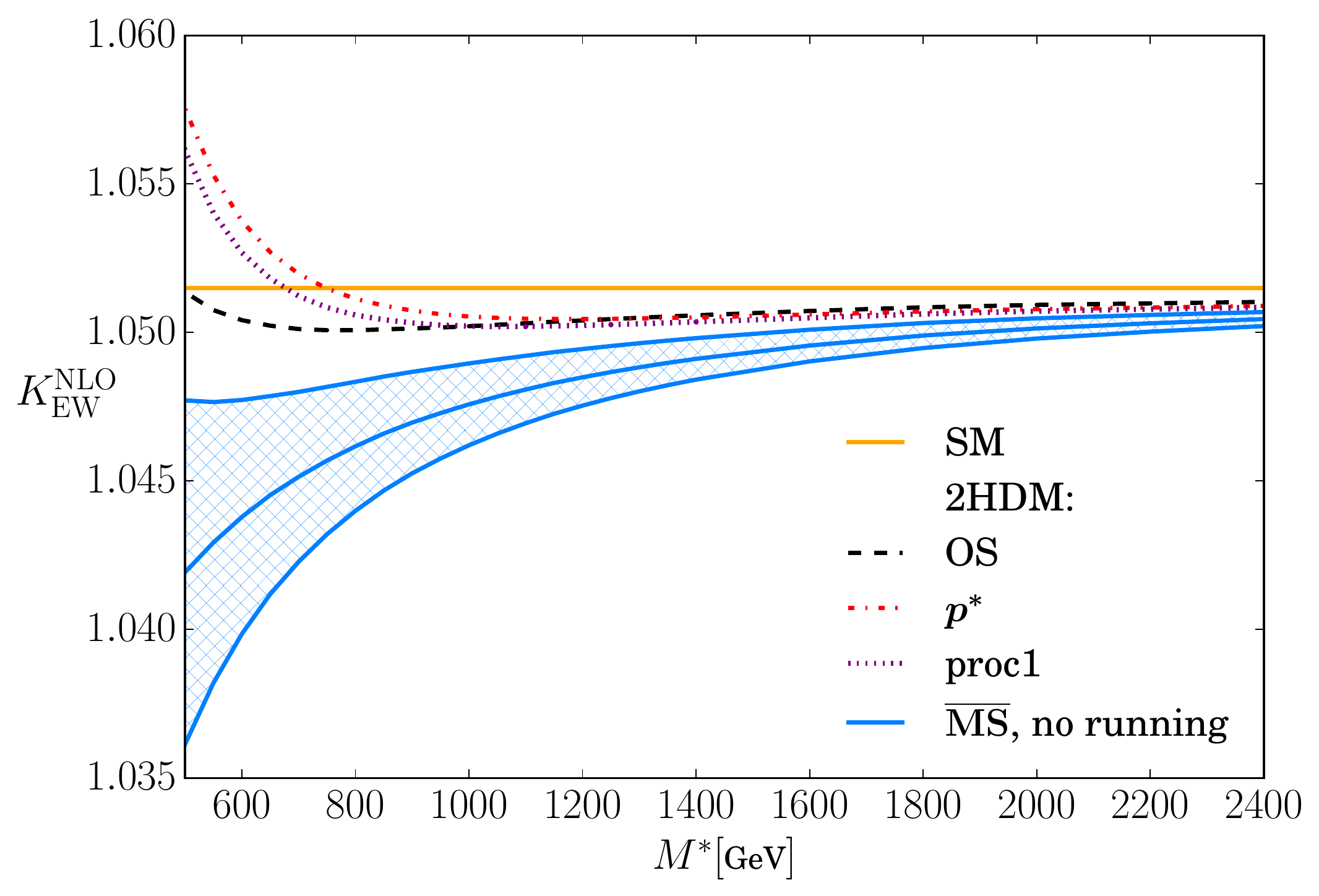}
  \end{center}
  \caption{Decoupling scenario for the process $g+g\to H_l$ as defined in
   Eqs.~(\ref{eq:decouplingparm}).
   \label{fig:decouplingggHl}}
\end{figure}

Another way to verify our implementation of the \THDM\ is to test the behaviour
of the NLO electroweak corrections in different limits.
In the decoupling limit introduced in the end of Section~\ref{sec:model},
all new Higgs-boson masses are much larger than the
electroweak scale, such that the new particles decouple from the SM. Therefore,
the NLO corrections in the \THDM\ should approach those of the SM.
Fig.~\ref{fig:decouplingggHl} shows a decoupling 
scenario for $g+g\to \hl$. The \THDM\ parameters have been chosen as
\begin{eqnarray}
  \Mhh=\Mha=\Mhc=\Msb=M^*,\quad
  \cab=0,\quad
  \tb=2,
\label{eq:decouplingparm}
\end{eqnarray}
where all masses are set equal to the scale of new physics $M^*$, 
because perturbativity requires the mass splitting to be smaller than 
$v^2/M^*$~\cite{Gunion:2002zf}.
At $M^*=2400$ GeV, the electroweak corrections in the \THDM\ have almost 
approached those of the SM. 
In addition, we observe that the scale dependence of the 
$\MSbar$-renormalized corrections shrinks, c.f.~Appendix~\ref{app:scale} 
for the analytic formula.
The proc2 renormalization has not been taken into account: 
since the LO of its defining process $\hh\to ZZ$ vanishes in the alignment 
limit, it cannot be expected to show proper decoupling behaviour.

Furthermore, several one-loop processes and all finite
NLO counterterms have been cross-checked against the
implementation in RECOLA2~\cite{Denner:2017vms} to
verify the implementation of the Feynman rules and the generation of one-loop
diagrams.

\section{Results and discussion\label{sec:results}}
In this section, we present the numerical results of phenomenologically
interesting scenarios for light and heavy Higgs-boson production in
gluon fusion for the CP-conserving \THDM.
First, we consider the benchmark points~(BP) in
Tabs.~\ref{tab:BPcab0} and \ref{tab:BPcab} in different renormalization
schemes. The BPs a-1 and b-1 are taken from Ref.~\cite{Baglio:2014nea} and they correspond to 
best-fit constraints on the triple Higgs couplings from the Higgs signal 
strengths.
All other BPs are from the LHC Higgs cross section working
group report~\cite{deFlorian:2016spz} and are sample points of phenomenologically interesting allowed 
scenarios compatible with the \THDM\ type II. Then, in order to analyze the influence
of the mass splitting on the
perturbative behaviour of the Higgs boson production processes,
we move to a benchmark scenario where we keep all new
Higgs boson masses fixed at 700~GeV except for the heavy Higgs boson mass
which is varied between 600~GeV and 800~GeV. This scenario agrees with current
experimental and theoretical exclusion limits~\cite{Aad:2015pla,Chowdhury:2017aav,Basler:2017nzu}.
\begin{table}[b]
  \centering
  \begin{tabular}{|c|c|c|c|c|c||c|c|c|}
    \hline
    \multirow{2}*{BP} & \multirow{2}*{$\ds{\Mhh\over\mathrm{GeV}}$}    
    & \multirow{2}*{$\ds{\Mha\over\mathrm{GeV}}$}
    & \multirow{2}*{$\ds{\Mhc\over\mathrm{GeV}}$}
    & \multirow{2}*{$\ds{m_{12}\over\mathrm{GeV}}$}
    & \multirow{2}*{$\tb$}
    & \multirow{2}*{$\ds{\Msb\over\mathrm{GeV}}$}
    & \multirow{2}*{$\ds{|\lambda_i^{\mbox{\tiny{max}}}|\over4\pi}$}
    & \multirow{2}*{$|\chh|^2$}\\
    &&&&&&&&\\
    \hline
    \hline
    $2_{1A}$ & 200.0 & 500.0 & 200.0 & 135.0 & 1.5 & 198.7 & 0.28 & 0.44 \\
    \hline
    $2_{1B}$ & 200.0 & 500.0 & 500.0 & 135.0 & 1.5 & 198.7 & 0.57 & 0.44 \\
    \hline
    $2_{1C}$ & 400.0 & 225.0 & 225.0 & 0.0 & 1.5 & 0.0 & 0.49 & 0.44 \\
    \hline
    $2_{1D}$ & 400.0 & 100.0 & 400.0 & 0.0 & 1.5 & 0.0 & 0.49 & 0.44 \\
    \hline
    $3_{A1}$ & 180.0 & 420.0 & 420.0 & 70.7 & 3.0 & 129.1 & 0.42 & 0.11 \\
    \hline
  \end{tabular}
  \caption{\THDM~benchmark points (BP) in the alignment limit,
    i.e. $\cab=0$, taken from Ref.~\cite{deFlorian:2016spz}. In the
    alignment limit, $|\chl|^2$ is always 1. \label{tab:BPcab0}}
\end{table}
\begin{table}
  \centering
  \begin{tabular}{|c|c|c|c|c|c|c||c|c|c|c|}
    \hline
    \multirow{2}*{BP} & \multirow{2}*{$\ds{\Mhh\over\mathrm{GeV}}$}    
    & \multirow{2}*{$\ds{\Mha\over\mathrm{GeV}}$}
    & \multirow{2}*{$\ds{\Mhc\over\mathrm{GeV}}$}
    & \multirow{2}*{$\ds{m_{12}\over\mathrm{GeV}}$}
    & \multirow{2}*{$\tb$}  & \multirow{2}*{$\cab$}
    & \multirow{2}*{$\ds{\Msb\over\mathrm{GeV}}$}
    & \multirow{2}*{$\ds{|\lambda_i^{\mbox{\tiny{max}}}|\over4\pi}$}
    & \multirow{2}*{$|\chl|^2$}
    & \multirow{2}*{$|\chh|^2$}\\
    &&&&&&&&&&\\
    \hline
    \hline
    a-1 & 700.0 & 700.0 & 670.0 & 424.3 & 1.5 & -0.091 & 624.5 & 0.16 & 0.87 & 0.57 \\
    \hline
    b-1 & 200.0 & 383.0 & 383.0 & 119.6 & 2.52 & -0.0346 & 204.2 & 0.30 & 0.97 & 0.19 \\
    \hline
    $2_{2A}$ & 500.0 & 500.0 & 500.0 & 187.1 & 7.0 & 0.28 & 500.0 & 0.64 & 1.00 & 0.02 \\
    \hline
    $3_{B1}$ & 200.0 & 420.0 & 420.0 & 77.8 & 3.0 & 0.3 & 142.0 & 0.44 & 1.11 & 0.0003 \\
    \hline
    $3_{B2}$ & 200.0 & 420.0 & 420.0 & 77.8 & 3.0 & 0.5 & 142.0 & 0.46 & 1.07 & 0.04 \\
    \hline
    $4_3$ & 263.7 & 6.3 & 308.3 & 52.3 & 1.9 & 0.14107 & 81.5 & 0.35 & 1.13 & 0.14 \\
    \hline
    $4_4$ & 227.1 & 24.7 & 226.8 & 58.4 & 1.8 & 0.14107 & 89.6 & 0.23 & 1.14 & 0.17 \\
    \hline
    $4_5$ & 210.2 & 63.06 & 333.5 & 69.2 & 2.4 & 0.71414 & 116.2 & 0.31 & 1.00 & 0.18 \\
    \hline
  \end{tabular}
  \caption{\THDM~benchmark points (BP) outside the alignment limit taken from
    Ref.~\cite{Baglio:2014nea} (a-1, b-1) and Ref.~\cite{deFlorian:2016spz}. \label{tab:BPcab}}
\end{table}
For the numerical evaluation we use the following set of SM input
parameters~\cite{Olive:2016xmw}:
\allowdisplaybreaks[0]
\begin{eqnarray}
    G_{F}&=&1.1663787\cdot10^{-5}~\GeV^{-2},\quad
    M_{W}=80.385~\GeV, \quad
    \Gamma_{W}=2.085~\GeV,\nonumber\\
    M_{Z}&=&91.1876~\GeV, \quad
    \Gamma_{Z}=2.4952~\GeV,\quad
    M_{t}=173.1~\GeV, \quad
    \Gamma_{t}=1.41~\GeV,\nonumber\\
    M_{h}&\equiv&\Mhl=125.09~\GeV.
\end{eqnarray}
\allowdisplaybreaks
For the renormalization of the $W$- and $Z$-boson masses we use the
complex mass scheme~\cite{Denner:1999gp,*Denner:2005fg,*Denner:2006ic}.
In contrast to Ref.~\cite{Denner:2016etu}, where the top-quark mass was
renormalized on-shell, here we also use the complex mass
scheme~\cite{Denner:1999gp,*Denner:2005fg,*Denner:2006ic} for the
top-quark mass.

\subsection*{Benchmark points}
According to the analysis of the LHC Higgs cross section working
group~\cite{deFlorian:2016spz},
the BPs in Tabs.~\ref{tab:BPcab0} and \ref{tab:BPcab} fulfill
constraints like perturbativity and vacuum stability. In Tabs.~\ref{tab:BPcab0} and
    \ref{tab:BPcab} we also provide the maximum of the couplings
$|\lambda_i^\mathrm{\tiny max}|/(4\pi)$ of the Higgs potential of Eq.~\eqref{eq:hpgb}. Compared to the SM value
$\lambda^{\mbox{\tiny SM}}/(4\pi)=0.02$, all $|\lambda_i^\mathrm{\tiny max}|/(4\pi)$
values are large.  
This can potentially lead
to large NLO corrections.

Tab.~\ref{tab:ggHlBPcab0Res} contains the NLO electroweak K-factors for
light Higgs-boson production in gluon fusion in different
renormalization schemes in the alignment limit. For this process, we
provide only the factor $\kew$, 
because the difference between $\kew$ and $\kbew$ is tiny -- at the
per mille level.
The \pOS, $p^*$ and the two
process-dependent renormalization schemes produce similar results and
mostly, the K-factors are comparable in size with the SM value
$K^\mathrm{\tiny SM}_{\ew}=1.051$. For the $\MSbar$-renormalized values,
we have chosen $\mu_0$ according to Eq.~(\ref{eq:mu0}),
different from the choice of the renormalization scale in
Ref.~\cite{Denner:2016etu}. The values for $\cab$ and
$\tb$ of the benchmark points in Tabs.~\ref{tab:BPcab0} and
\ref{tab:BPcab} are defined at the scale $\mu_0$. 
The uncertainty is obtained as described in
Section~\ref{sec:renormalization} by taking one half and twice the
central value $\mu_0$ and running the parameters $\cab$ and $\tb$ to the
corresponding scales. The results in the $\MSbar$ scheme are different
    from the other results and the benchmark points $2_{1A}$ and
    $2_{1B}$ show a large scale dependence as already analyzed in
    Ref.~\cite{Denner:2016etu}, where no running of the parameters was
    considered yet. The running results in an enhanced
    scale dependence for these two benchmark points.
\begin{table}[t]
  \centering
  \begin{tabular}{|c|c|c|c|c|c|Sc|}
    \hline
    \multirow{2}*{BP}  &
    \multirow{2}*{$\ds{\hat{\sigma}^{\mbox{\tiny{LO}}}\over\mathrm{fb}}$} &
    \multirow{2}*{$\ds K_{\ew}^{\mbox{\tiny{OS}}}$}     &
    \multirow{2}*{$\ds K_{\ew}^{p^*}$}                  &
    \multirow{2}*{$\ds K_{\ew}^{\mbox{\tiny{proc1}}}$}  &
    \multirow{2}*{$\ds K_{\ew}^{\mbox{\tiny{proc2}}}$}  &
    \multirow{2}*{$\ds K_{\ew}^{\overline{\mbox{\tiny{MS}}}}$}\\
    &&&&&&\\
    \hline
    \hline
    $2_{1A}$  &  51.856  &  1.053 & 1.063 & 1.101 & 1.037  &  0.994 $\substack{ -0.030\\+0.342 }$ \\
    \hline
    $2_{1B}$  &  51.856  &  1.038 & 1.048 & 1.044 & 1.022  &  0.930 $\substack{ +0.066\\+2.389 }$ \\
    \hline
    $2_{1C}$  &  51.856  &  1.043 & 1.044 & 1.099 & 1.055  &  1.126 $\substack{ -0.001\\-0.007 }$ \\
    \hline
    $2_{1D}$  &  51.856  &  1.029 & 1.035 & 1.042 & 1.029  &  1.145 $\substack{ -0.012\\-0.015 }$ \\
    \hline
    $3_{A1}$  &  51.856  &  1.041 & 1.040 & 1.045 & 1.041  &  1.118 $\substack{ -0.042\\-0.010 }$ \\
	\hline
  \end{tabular}
  \caption{The NLO electroweak K-factors for
    the process $g+g\to \hl$ for the \THDM~benchmark points (BP) in the
    alignment limit are shown. They are defined in Tab.~\ref{tab:BPcab0}.
    The renormalization scale has been set to $\mu_0$ as defined in Eq.~\eqref{eq:mu0} for the $\MSbar$
    scheme.  The lower (upper) value of the $\MSbar$ result corresponds
    to the change when
     taking  $\mu_0/2$ ($2\mu_0$) as renormalization scale. 
\label{tab:ggHlBPcab0Res}}
\end{table}

For the remaining benchmark points with $\cab\ne0$, the K-factors of the electroweak
corrections to the process $g+g\to\hl$ are shown likewise in Tab.~\ref{tab:ggHlBPcabRes}.
Again, the K-factors of the \pOS, $p^*$ and the two process-dependent
renormalization schemes
are close to the SM, while the $\MSbar$ scheme leads to different results.
Nevertheless, the electroweak corrections are always moderate in size,
i.e.~they are mostly
below 5\% compared to the LO production cross section.
\begin{table}[!b]
  \centering
  \begin{tabular}{|c|c|c|c|c|c|Sc|}
    \hline
    \multirow{2}*{BP}  &
    \multirow{2}*{$\ds{\hat{\sigma}^{\mbox{\tiny{LO}}}\over\mathrm{fb}}$} &
    \multirow{2}*{$\ds K_{\ew}^{\mbox{\tiny{OS}}}$}     &
    \multirow{2}*{$\ds K_{\ew}^{p^*}$}                  &
    \multirow{2}*{$\ds K_{\ew}^{\mbox{\tiny{proc1}}}$}  &
    \multirow{2}*{$\ds K_{\ew}^{\mbox{\tiny{proc2}}}$}  &
    \multirow{2}*{$\ds K_{\ew}^{\overline{\mbox{\tiny{MS}}}}$}\\
    &&&&&&\\
    \hline
    \hline
    a-1  &  45.352  & 1.043 & 1.047 & 1.048 & 1.057  &  0.962 $\substack{ +0.042\\+2.634 }$ \\
    \hline
    b-1  &  50.381  &  1.048 & 1.045 & 1.054 & 1.040  &  0.995 $\substack{ +0.002\\+0.175 }$ \\
    \hline
    $2_{2A}$  &  51.856  &  1.017 & 1.018 & 1.015 & 1.017  &  1.006 $\substack{ -0.871\\\mbox{--} }$ \\
    \hline
    $3_{B1}$  &  57.601  &  1.039 & 1.038 & 1.039 & 1.032  &  1.072 $\substack{ -0.152\\+0.079 }$ \\
    \hline
    $3_{B2}$  &  55.302  &  1.037 & 1.036 & 1.035 & 1.034  &  0.917 $\substack{ -0.054\\+0.120 }$ \\
    \hline
    $4_3$  &  58.733  &  1.042 & 1.043 & 1.038 & 1.036  &  1.126 $\substack{ -0.022\\+0.030 }$ \\
    \hline
    $4_4$  &  59.189  &  1.043 & 1.044 & 1.038 & 1.034  &  1.103 $\substack{ +0.002\\+0.002 }$ \\
    \hline
    $4_5$  &  51.603  &  1.037 & 1.036 & 1.026 & 1.070  &  1.045 $\substack{ +0.111\\-0.075 }$ \\
    \hline
  \end{tabular}
  \caption{The NLO electroweak K-factors for the process
    $g+g\to \hl$ for the \THDM~benchmark points~(BP) that are not in the
    alignment limit are presented. They are defined in Tab.~\ref{tab:BPcab}.
    The renormalization scale has been set to $\mu_0$ as defined in Eq.~\eqref{eq:mu0} for the $\MSbar$
    scheme. The lower (upper) value of the $\MSbar$ result corresponds
    to the change when
    taking $\mu_0/2$ ($2\mu_0$)  as renormalization scale.
    The lower $\MSbar$ uncertainty for BP
        $2_{2A}$ is not shown, since the RGEs~(\ref{eq:running}) do not allow for a
        stable solution, see also Fig.~\ref{fig:lambdas_BPs} of Appendix~\ref{app:scale_plots}.
    \label{tab:ggHlBPcabRes}}
\end{table}
\begin{table}[t]
  \centering
  \begin{tabular}{|c|c|c|c|c|c|Sc|}
    \hline
    \multirow{2}*{BP}  &
    \multirow{2}*{$\ds{\hat{\sigma}^{\mbox{\tiny{LO}}}\over\mathrm{fb}}$} &
    \multirow{2}*{$\ds \kb_{\ew}^{\mbox{\tiny{OS}}}$}     &
    \multirow{2}*{$\ds \kb_{\ew}^{p^*}$}                  &
    \multirow{2}*{$\ds \kb_{\ew}^{\mbox{\tiny{proc1}}}$}  &
    \multirow{2}*{$\ds \kb_{\ew}^{\mbox{\tiny{proc2}}}$}  &
    \multirow{2}*{$\ds \kb_{\ew}^{\overline{\mbox{\tiny{MS}}}}$}\\
    &&&&&&\\
    \hline
    \hline
    \multirow{2}*{     $2_{1A}$ } & \multirow{2}*{ 25.737 } &  0.486 & 0.492 & 0.426 & 0.533  &  0.655 $\substack{ +0.197\\-0.451 }$ \\
    & & (0.360) & (0.369) & (0.276) & (0.420) & $\left(0.590 \substack{ +0.229\\-0.733 }\right)$ \\
    \hline
    \multirow{2}*{     $2_{1B}$ } & \multirow{2}*{ 25.737 } &  0.177 & 0.178 & 0.166 & 0.183  &  0.257 $\substack{ -0.033\\-0.060 }$ \\
    & & (-0.765) & (-0.756) & (-0.823) & (-0.773) & $\left(-0.387 \substack{ -0.181\\-1.336 }\right)$ \\
    \hline
    \multirow{2}*{     $2_{1C}$ } & \multirow{2}*{ 69.019 } &  0.958 & 0.950 & 0.822 & 0.904  &  0.822 $\substack{ +0.176\\-0.088 }$ \\
    & & (0.939) & (0.931) & (0.780) & (0.879) & $\left(0.793 \substack{ +0.181\\-0.097 }\right)$ \\
    \hline
    \multirow{2}*{     $2_{1D}$ } & \multirow{2}*{ 69.019 } &  0.854 & 0.840 & 0.803 & 0.848  &  0.693 $\substack{ +0.244\\-0.095 }$ \\
    & & (0.845) & (0.831) & (0.788) & (0.816) & $\left(0.662 \substack{ +0.269\\-0.117 }\right)$ \\
    \hline
    \multirow{2}*{     $3_{A1}$ } & \multirow{2}*{ 6.205 } &  0.581 & 0.580 & 0.486 & 0.550  &  0.336 $\substack{ +0.667\\-0.100 }$ \\
    & & (0.305) & (0.303) & (0.210) & (0.245) & $\left(-0.150 \substack{ -0.223\\-0.121 }\right)$ \\
    \hline
  \end{tabular}
  \caption{The NLO electroweak K-factors for
    the process $g+g\to \hh$ for the \THDM~benchmark points (BP) in the
    alignment limit are shown. They are defined in Tab.~\ref{tab:BPcab0}. Both $\kbew$ (first value) and $\kew$
    (in parentheses) are shown.  The renormalization scale has been set to $\mu_0$ as defined in Eq.~\eqref{eq:mu0}
    for the $\MSbar$ scheme. The lower (upper) value of the $\MSbar$ result corresponds
    to the change when taking $\mu_0/2$ ($2\mu_0$) as renormalization scale. 
    \label{tab:ggHhBPcab0Res}}
\end{table}
We now turn to heavy, neutral Higgs-boson production $g+g\to \hh$.
There are two aspects that we have to consider when looking at this
process. On the one hand, the LO production cross sections in
Tabs.~\ref{tab:ggHhBPcab0Res} and~\ref{tab:ggHhBPcabRes}
depend on the heavy Higgs-boson mass as well as on the coefficient $|\chh|^2$.
On the other hand, the coefficients $|\chh|^2$ can be close to zero.
This is different from the light Higgs-boson production case, where
for all BPs the same light Higgs-boson mass enters,
and where the coefficients $|\chl|^2$,
as given in Tab.~\ref{tab:BPcab}, are always close to 1 or even equal to 1,
as it is the case in the alignment limit.
For these reasons, the LO cross sections of the heavy Higgs-boson production
can be very small and in particular comparable in size to the NLO contribution.
Then, further terms of the perturbative expansion of the cross section
are required
\begin{equation}
\label{eq:ANNLO}
\begin{aligned}
  |A_{\hh}^{\mbox{\tiny NNLO,EW}}|^2 & =
  |A_{\hh}^{(1)}+A_{\hh}^{(2)}+A_{\hh}^{(3)}|^2\\
  & =
  |A_{\hh}^{(1)}|^2
  +A_{\hh}^{(1)}A_{\hh}^{(2)*}+A_{\hh}^{(1)*}A_{\hh}^{(2)}
  +|A_{\hh}^{(2)}|^2\\
  & \phantom{=} + A_{\hh}^{(1)}A_{\hh}^{(3)*}+A_{\hh}^{(1)*}A_{\hh}^{(3)}
  + \mathcal{O}\left((\Gf\Mw^2)^3\right),\\
\end{aligned}
\end{equation}
where $A_{\hh}^{(3)}$ is the three-loop amplitude. In Eq.~(\ref{eq:ANNLO}),
$|A_{\hh}^{(1)}|^2$ is the LO cross section up to a global
factor. Likewise, $A_{\hh}^{(1)}A_{\hh}^{(2)*}+A_{\hh}^{(1)*}A_{\hh}^{(2)}$
corresponds to the NLO contribution,
while $|A_{\hh}^{(2)}|^2$ and $A_{\hh}^{(1)}A_{\hh}^{(3)*}+A_{\hh}^{(1)*}A_{\hh}^{(3)}$
enter at NNLO. If $|A_{\hh}^{(1)}|^2$ is very small, $|A_{\hh}^{(2)}|^2$ can be of
the same size as the NLO contribution,
while $A_{\hh}^{(1)}A_{\hh}^{(3)*}+A_{\hh}^{(1)*}A_{\hh}^{(3)}$ can be expected
to be small again. Therefore, $|A_{\hh}^{(2)}|^2$ should not be neglected.
Furthermore, the imaginary part of the LO amplitude can be similar in magnitude or
even larger than the real part when the heavy Higgs-boson mass becomes
larger than twice the top-quark mass. This can lead to cancellations between
real and imaginary contributions in the NLO term.

In both cases, there can be a considerable difference between $\kew$ and
$\kbew$ as defined in Eqs.~\eqref{eq:Kfactor} and~\eqref{eq:Kbfactor}.
Since $\kbew$ contains the additional contribution,
it seems to be more appropriate to quantify the electroweak corrections
to heavy Higgs-boson production.
In order to point out BPs where this difference occurs, the
$\kew$~are shown in parentheses in Tabs.~\ref{tab:ggHhBPcab0Res}
and~\ref{tab:ggHhBPcabRes} even though they can take an
unphysical, negative value for some benchmark points, e.g. for $2_{1B}$ and $3_{B1}$.
\begin{table}[t]
  \centering
  \begin{tabular}{|c|c|c|c|c|c|c|Sc|}
    \hline
    \multirow{2}*{BP}  &
    \multirow{2}*{$\ds{\hat{\sigma}^{\mbox{\tiny{LO}}}\over\mathrm{fb}}$} &
    \multirow{2}*{$\ds \kb_{\ew}^{\mbox{\tiny{OS}}}$}     &
    \multirow{2}*{$\ds \kb_{\ew}^{p^*}$}                  &
    \multirow{2}*{$\ds \kb_{\ew}^{\mbox{\tiny{proc1}}}$}  &
    \multirow{2}*{$\ds \kb_{\ew}^{\mbox{\tiny{proc2}}}$}  &
    \multirow{2}*{$\ds \kb_{\ew}^{\overline{\mbox{\tiny{MS}}}}$}\\
    &&&&&&\\
    \hline
    \hline
    \multirow{2}*{     a-1 } & \multirow{2}*{ 45.488 } &  1.145 & 1.161 & 1.121 & 1.084  &  1.401 $\substack{ +0.253\\-1.385 }$ \\
    & & (1.031) & (1.048) & (1.011) & (0.997) & $\left(1.270 \substack{ +0.200\\-2.114 }\right)$ \\
    \hline
    \multirow{2}*{     b-1 } & \multirow{2}*{ 10.767 } &  0.696 & 0.693 & 0.628 & 0.701  &  0.981 $\substack{ +0.154\\-0.574 }$ \\
    & & (0.569) & (0.566) & (0.499) & (0.572) & $\left(0.898 \substack{ +0.140\\-0.713 }\right)$ \\
    \hline
    \multirow{2}*{     $2_{2A}$ } & \multirow{2}*{ 2.866 } &  7.504 & 7.456 & 7.517 & 6.989  &  4.030 $\substack{ +0.758\\\mbox{--} }$ \\
    & & (2.798) & (2.773) & (2.812) & (2.738) & $\left(-0.273 \substack{ +3.301\\\mbox{--} }\right)$ \\
    \hline
    \multirow{2}*{     $3_{B1}$ } & \multirow{2}*{ 0.019 } &  27.00 & 27.09 & 326.3 & 37.35  &  286.5 $\substack{ -284.9\\-285.5 }$ \\
    & & (-2.714) & (-2.821) & (-24.62) & (-2.739) & $\left(-33.26 \substack{ +34.70\\+30.66 }\right)$ \\
    \hline
    \multirow{2}*{     $3_{B2}$ } & \multirow{2}*{ 2.586 } &  1.019 & 1.031 & 1.005 & 1.042  &  7.491 $\substack{ -6.210\\+23.13 }$ \\
    & & (0.933) & (0.945) & (0.828) & (0.850) & $\left(4.443 \substack{ -3.181\\+5.347 }\right)$ \\
    \hline
    \multirow{2}*{     $4_3$ } & \multirow{2}*{ 9.828 } &  0.945 & 0.941 & 0.979 & 1.151  &  0.580 $\substack{ +0.531\\-0.171 }$ \\
    & & (0.887) & (0.883) & (0.910) & (0.925) & $\left(0.450 \substack{ +0.574\\-0.237 }\right)$ \\
    \hline
    \multirow{2}*{     $4_4$ } & \multirow{2}*{ 10.271 } &  1.028 & 1.024 & 1.067 & 1.216  &  0.807 $\substack{ +0.322\\-0.190 }$ \\
    & & (0.977) & (0.973) & (1.006) & (1.032) & $\left(0.740 \substack{ +0.326\\-0.224 }\right)$ \\
    \hline
    \multirow{2}*{     $4_5$ } & \multirow{2}*{ 10.552 } &  0.794 & 0.799 & 0.844 & 0.749  &  1.096 $\substack{ -0.402\\+1.155 }$ \\
    & & (0.782) & (0.788) & (0.838) & (0.591) & $\left(1.094 \substack{ -0.429\\+0.905 }\right)$ \\
    \hline
  \end{tabular}
  \caption{The NLO electroweak K-factors for the process
    $g+g\to \hh$ for the \THDM~benchmark points (BP) that are not in the
    alignment limit are shown. They are defined in Tab.~\ref{tab:BPcab}.
    The renormalization scale has been set to $\mu_0$ as defined in Eq.~\eqref{eq:mu0}
    for the $\MSbar$ scheme. Both $\kbew$ (first value) and $\kew$
    (in parentheses) are shown. The lower (upper) value of the $\MSbar$ 
    result corresponds to the change when taking $\mu_0/2$ ($2\mu_0$)  as 
    renormalization scale. The lower $\MSbar$ uncertainty for BP
        $2_{2A}$ is not shown, since the RGEs~(\ref{eq:running}) do not allow for a
        stable solution, see also Fig.~\ref{fig:lambdas_BPs} of Appendix~\ref{app:scale_plots}.
    \label{tab:ggHhBPcabRes}}
\end{table}
Different from the sister process $g+g\to\hl$, many K-factors in
Tabs.~\ref{tab:ggHhBPcab0Res} and~\ref{tab:ggHhBPcabRes} are far from
one. This is at least partly due to the different LO couplings between
the $t-H_{l}-t$ and $t-H_{h}-t$ interactions. While the $|\chl|^2$
in Tabs.~\ref{tab:BPcab0} and~\ref{tab:BPcab} are generally one or close to one,
the $|\chh|^2$ are at most 0.57. In addition, the decoupling of the
new Higgs sector from the SM can play a role for BP a-1 and $2_{2A}$. While
decoupling imposes the restriction that SM processes may not receive large
corrections from the new sector, this is not the case for non-SM processes like
heavy Higgs-boson production.

In the following, we have a closer look at the benchmark points
$2_{1B}$, $2_{2A}$ and $3_{B1}$, which have
large electroweak NLO corrections and a large difference between $\kbew$ and $\kew$ for heavy Higgs-boson production.
Benchmark point $2_{1B}$ does not only show very large
corrections of more than -80\%, but also the difference between $\kbew$ and $\kew$
is of the same order of magnitude. Hence, $|A_{\hh}^{(1)}|$ and
$|A_{\hh}^{(2)}|$ must be similar in size, which may partially be due to the
large value of
$|\lambda_i^\mathrm{\tiny max}|/(4\pi)=0.57$. The same is true for benchmark point
$2_{2A}$. It has a large value of $|\lambda_i^\mathrm{\tiny max}|/(4\pi)=0.64$, 
and in addition, its LO cross section is quite small, such that the two-loop diagrams
yield the true LO contribution. In order to verify whether the NNLO contribution
is small in these scenarios, would, however, require the calculation
of the three-loop contributions, which is not within reach in the near future.

The benchmark point $3_{B1}$ has an even
smaller LO cross section due to the almost vanishing coefficient $|\chh|^2=0.0003$.
Therefore, the extremely large K-factor of more than 27 can be understood,
since the true LO contribution is given by the two-loop diagrams. However, since the
coefficients in front of the counterterms of the mixing angles are large for this BP
and the small LO does not factorize, we observe large differences between the results
in different renormalization schemes. This leads to a large dependence on the choice
of the renormalization scheme, and thus to large theoretical uncertainties on
the production cross section. In particular, this shows that a small coupling of the
heavy Higgs boson to top quarks does not automatically lead to a small cross section
when higher order contributions are considered.

Finally, there
are benchmark points that have moderate NLO corrections on the one hand,
but a relatively large difference between $\kbew$ and $\kew$ on the other
hand. We have investigated this difference further and found that it is
caused by cancellations between the real and imaginary contributions in the
term $A_{\hh}^{(1)}A_{\hh}^{(2)*}+A_{\hh}^{(1)*}A_{\hh}^{(2)}$.
Especially benchmark point a-1
exhibits this feature, but it can also be observed in BP $3_{B2}$ and $4_4$.

As a measure of the scale dependence in the $\MSbar$ scheme, we can look
at the size of the coefficients of the scale-dependent logarithms in the
percentage correction presented in Tabs.~\ref{tab:BPsscalecab0} and~\ref{tab:BPsscalecabn0} 
of Appendix~\ref{app:scale}.
In general, they can become very small or even vanish for certain choices
of the parameters. In particular, in the alignment limit with $\Msb=0$,
the scale dependent logarithmic term for $g+g\to\hl$ can be obtained from
    Eq.~\eqref{eq:scaledep_gghl} of Appendix~\ref{app:scale} 
and becomes very simple
\begin{equation}
\label{eq:deltascaledep}
\delta^{\mbox{\scriptsize{NLO}},\mu\mathrm{{-}dep.}}_{\mbox{\scriptsize EW}, \Msb=0}
=
{3\,\Gf\*\sqrt{2}\,\Mhl^2\over8\*\pi^2\*\tb^2\*(\Mhh^2-\Mhl^2)}\ln\!\left({\mu^2\over\Mhl^2}\right)
\left[ (1-\tb^2)\Mhh^2 + 2\mt^2 \right],
\end{equation}
which even vanishes for the special case $\Mhh^2= 2\mt^2/(\tb^2-1)$. This scenario is almost realized
for BP $2_{1C}$ and $2_{1D}$, and hence, these benchmark points exhibit only a small scale
dependence in Tab.~\ref{tab:ggHlBPcab0Res}.

In addition, we observe that the $\MSbar$ corrections for light Higgs-boson production
are usually moderate, while for heavy Higgs-boson production, this scheme often
leads to K-factors far from one
in addition to a very strong scale dependence. This can,
again, partly be traced back to small LO couplings between the heavy Higgs boson
and the top quarks.

\subsection*{Benchmark scenarios at $\boldsymbol{M^*=700}$~GeV\label{sec:Mstar}}
\begin{figure}[!h]
  \begin{center}
    \includegraphics[width=11cm,bb=0 0 782 432]{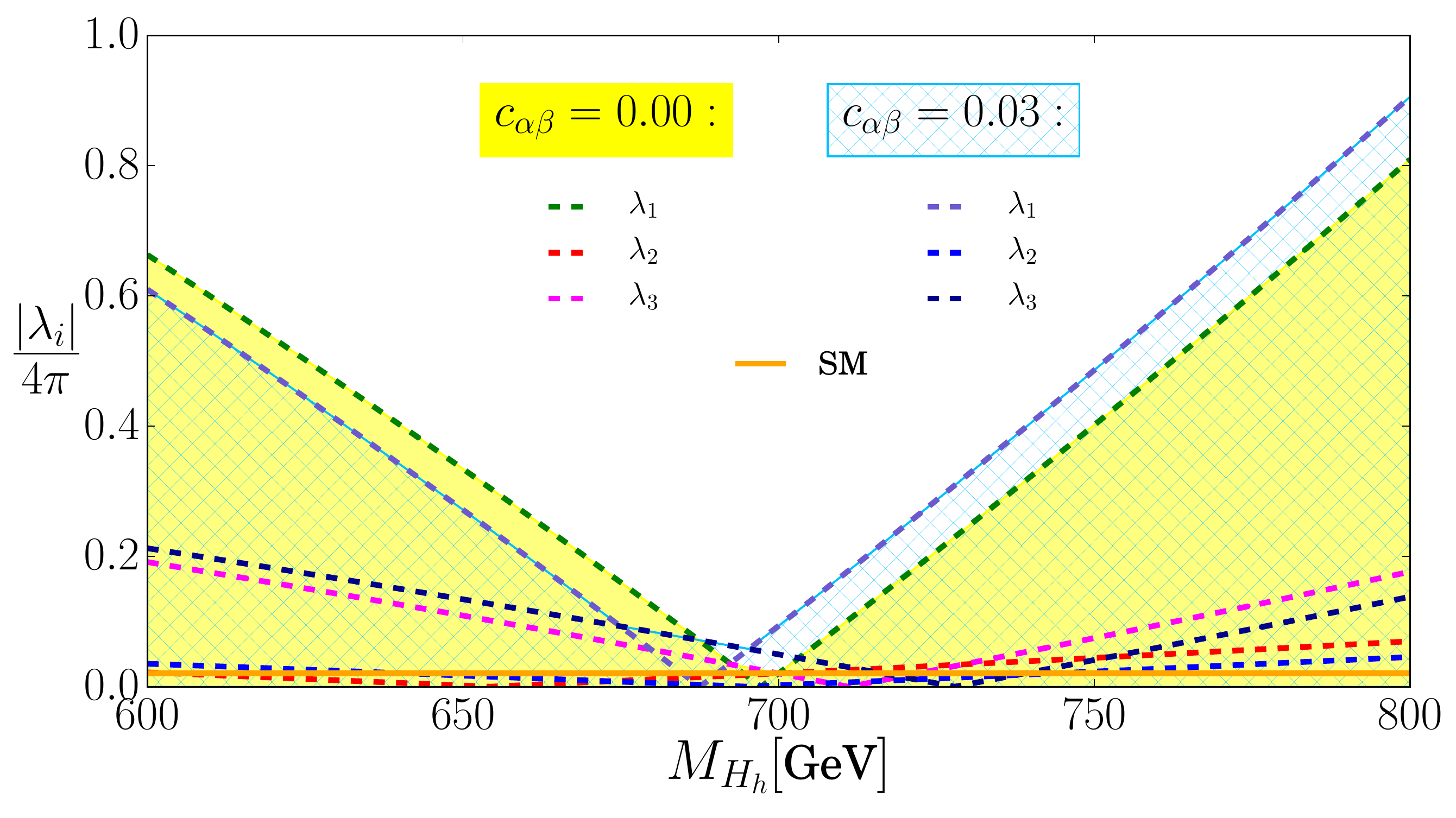}
  \end{center}
  \caption{The range of $|\lambda_i|/(4\pi)$ for our benchmark
    scenarios compared to the SM value
   $\lambda^{\mbox{\tiny{SM}}}/(4\pi)=0.02$ (solid,
        orange line). For the two different values of $\cab$ the
        individual curves for those $|\lambda_i|/(4\pi)$ that are
        different from zero are also shown (dashed lines). The explicit formulae for
        the parameters $\lambda_i$ are given in
        Eqs.~(\ref{eq:lambda1})-(\ref{eq:lambda5}) of
        Appendix~\ref{app:scale_plots}. The values shown here are for
        the non-$\MSbar$ schemes and thus no running is taken into
        account.\label{fig:lambda}}
\end{figure}

Next, we consider light and heavy Higgs-boson production in two benchmark
scenarios at a moderately heavy mass scale $M^*$
\begin{equation}
\label{eq:Msscenario}
  \Mha=\Mhc=\Msb=M^*=700\,\text{GeV},\quad\tb=2,\quad\Mhh=600\ldots 800\,\text{GeV}.
\end{equation}
The first scenario uses the alignment limit and the second scenario sets
$\cab=0.03$. As we see in Fig.~\ref{fig:lambda}, both scenarios fulfill
the perturbativity restriction $|\lambda_i|/(4\pi)<1$ ($i=1,...,5$). 
In addition, they respect the experimental constraints considered in 
Refs.~\cite{Aad:2015pla,Chowdhury:2017aav}.
In both scenarios we vary the heavy, scalar Higgs-boson mass between
$600$ and $800$~GeV. The size of the couplings $\lambda_i$
is more sensitive to these variations compared to variations in the
other Higgs-boson masses $\Mha$, $\Mhc$ (compare
    Fig.~\ref{fig:lambda} to Fig.~\ref{fig:scan-lambdas-app} in Appendix~\ref{app:scale_plots}).
For that reason we will present the NLO corrections to
Higgs-boson production and also the scale dependence in the $\MSbar$
renormalization scheme as a function of the heavy Higgs-boson mass in the following.
\begin{figure}[!h]
  \begin{center}
    \begin{minipage}{13cm}
      \begin{center}
                \includegraphics[width=12.85cm,bb=0 0 507 379]{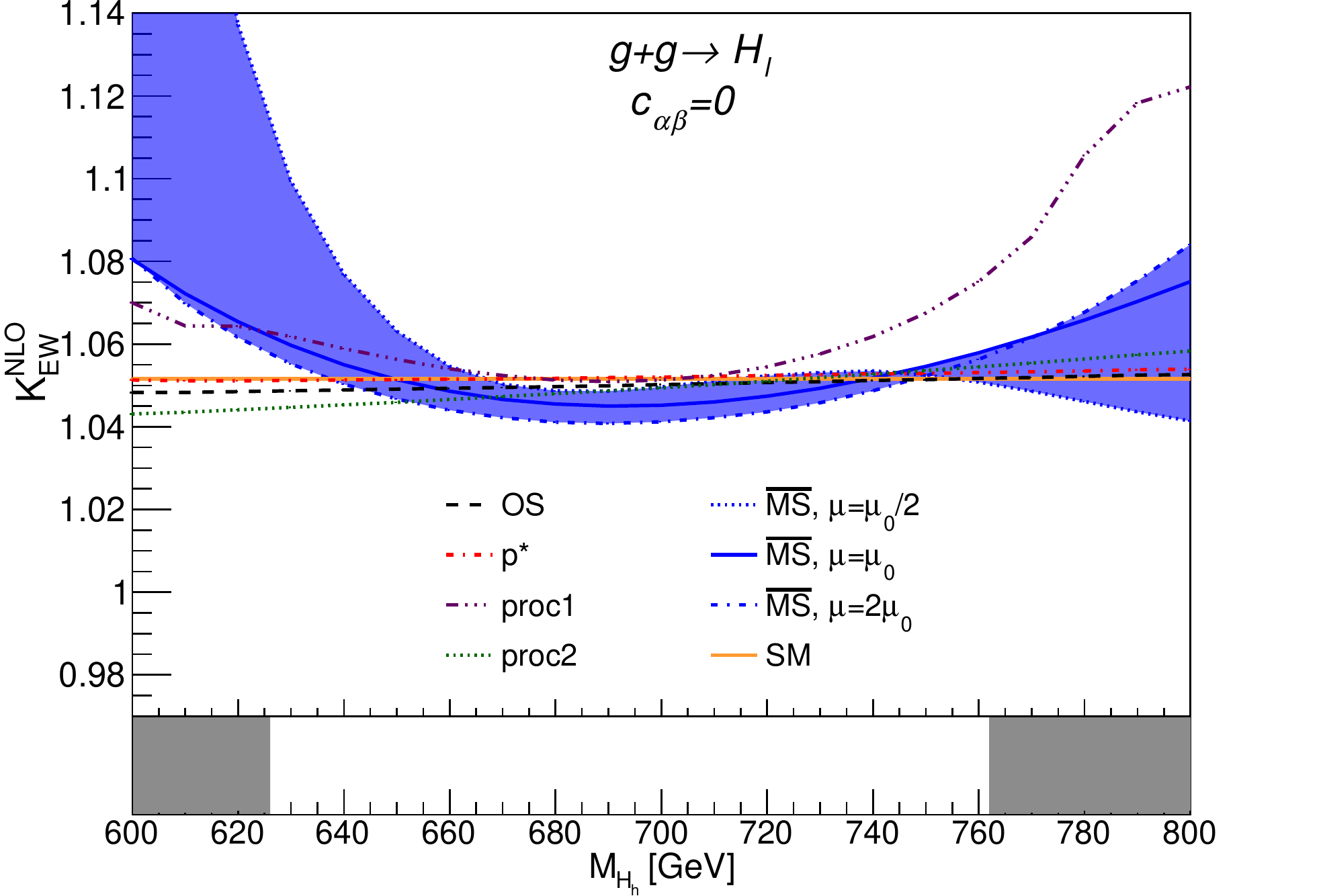}
      \end{center}
    \end{minipage}\\[0.25cm]
    \begin{minipage}{13cm}
      \begin{center}
                \includegraphics[width=12.85cm,bb=0 0 507 379]{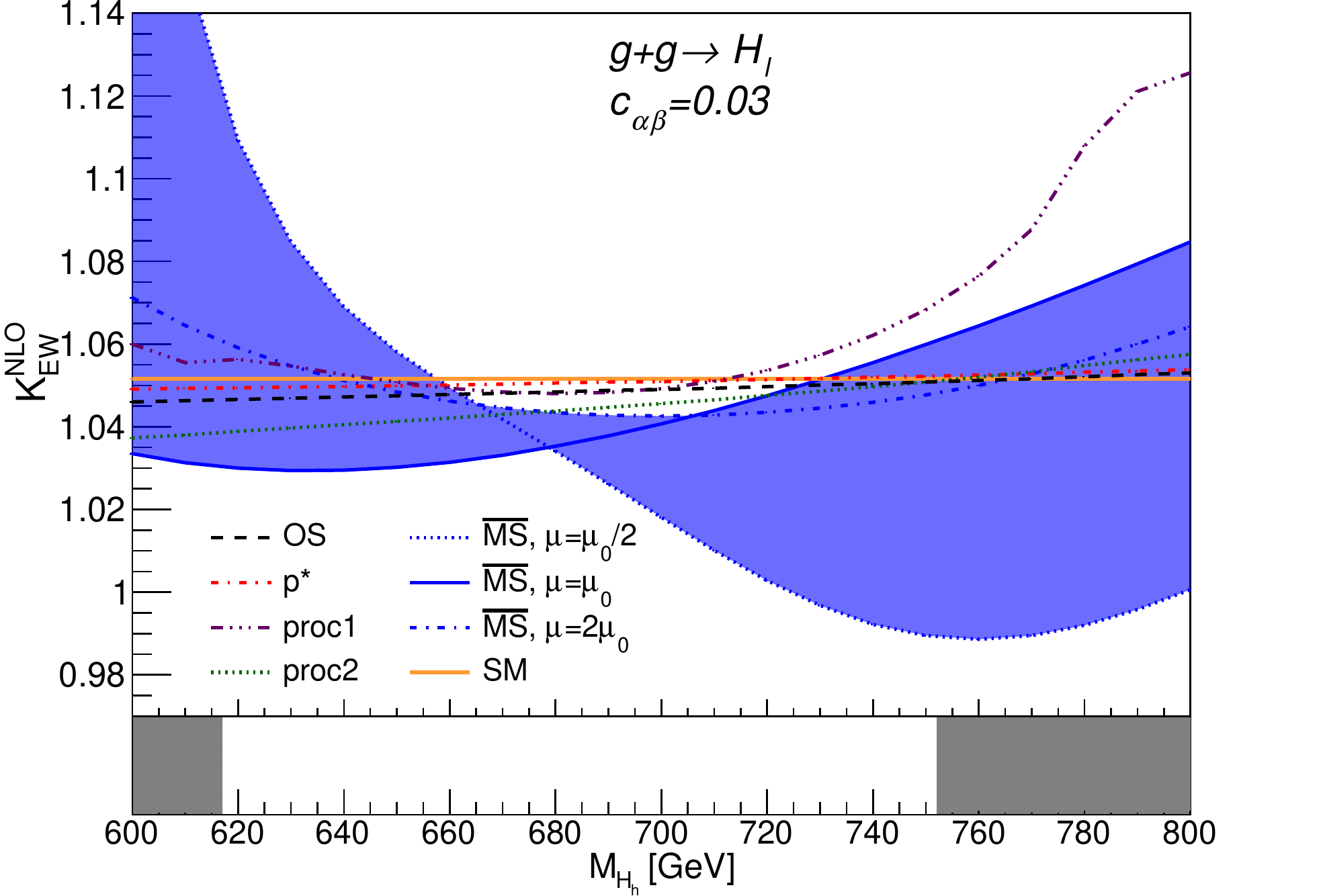}
      \end{center}
    \end{minipage}
  \end{center}
  \caption{Percentage correction of the process $g+g\to\hl$ in various
    schemes. The grey shaded band denotes the region where at
        least one of the couplings $|\lambda_i|/(4\pi)$ becomes larger than
        0.5 and where  one slowly starts to enter in the non-perturbative
        regime.\label{fig:scaleggHl}}
\end{figure}
Fig.~\ref{fig:scaleggHl} shows the K-factor of the NLO electroweak
corrections for the process $g+g\to\hl$ in the \THDM\ as a function of
the heavy, neutral Higgs-boson mass for the different renormalization
schemes of the mixing angles $\alpha$ and $\beta$.  We consider the
\pOS, $p^*$, two process-dependent and the $\MSbar$ scheme.  The grey
shaded band shows
the region where at least one of the couplings $|\lambda_i|/(4\pi)$ of
Eq.~(\ref{eq:hpgb}) becomes larger than 0.5. The region of variation of the
corresponding $|\lambda_i|/(4\pi)$ values is displayed in
Fig.~\ref{fig:lambda}. These values are not valid for the $\MSbar$
scheme, where they adopt different values depending on the choice of the
renormalization scale. As a result of this, the region of perturbativity generally
looks different for the $\MSbar$ renormalized results.
In the $\MSbar$ scheme the corrections are renormalization-scale
dependent and we choose the central renormalization scale $\mu_0$ 
according to Eq.~\eqref{eq:mu0}. In order to estimate the impact of the scale
dependence on the size of the NLO EW corrections, we vary the scale
between $\mu_0/2$ and $2\mu_0$. A more detailed analysis is discussed in
Appendix~\ref{app:scaledep}.
In the alignment limit presented in the first plot of
Fig.~\ref{fig:scaleggHl}, the different renormalization schemes agree
quite well for $|\lambda_i|/(4\pi)<0.5$.  The curves for $\MSbar$
renormalization (blue) and the renormalization scheme proc1 (purple,
dashed-dotted) show the biggest deviation from the SM, 
especially when entering the non-perturbative region.   The corrections of the \pOS, $p^*$ and proc2 renormalization
schemes are very close to each other for these scenarios, even for large
$\lambda_i$.
For $\cab=0.03$ presented in the second plot of
Fig.~\ref{fig:scaleggHl}, the behaviour of the \pOS, $p^*$, proc1 and
proc2 scheme is very similar to the case of the alignment limit~(upper
plot), while the $\MSbar$ results show a much larger scale variation.

\begin{figure}[!h]
  \begin{center}
    \begin{minipage}{13cm}
      \begin{center}
        \includegraphics[width=12.85cm,bb=0 0 509 379]{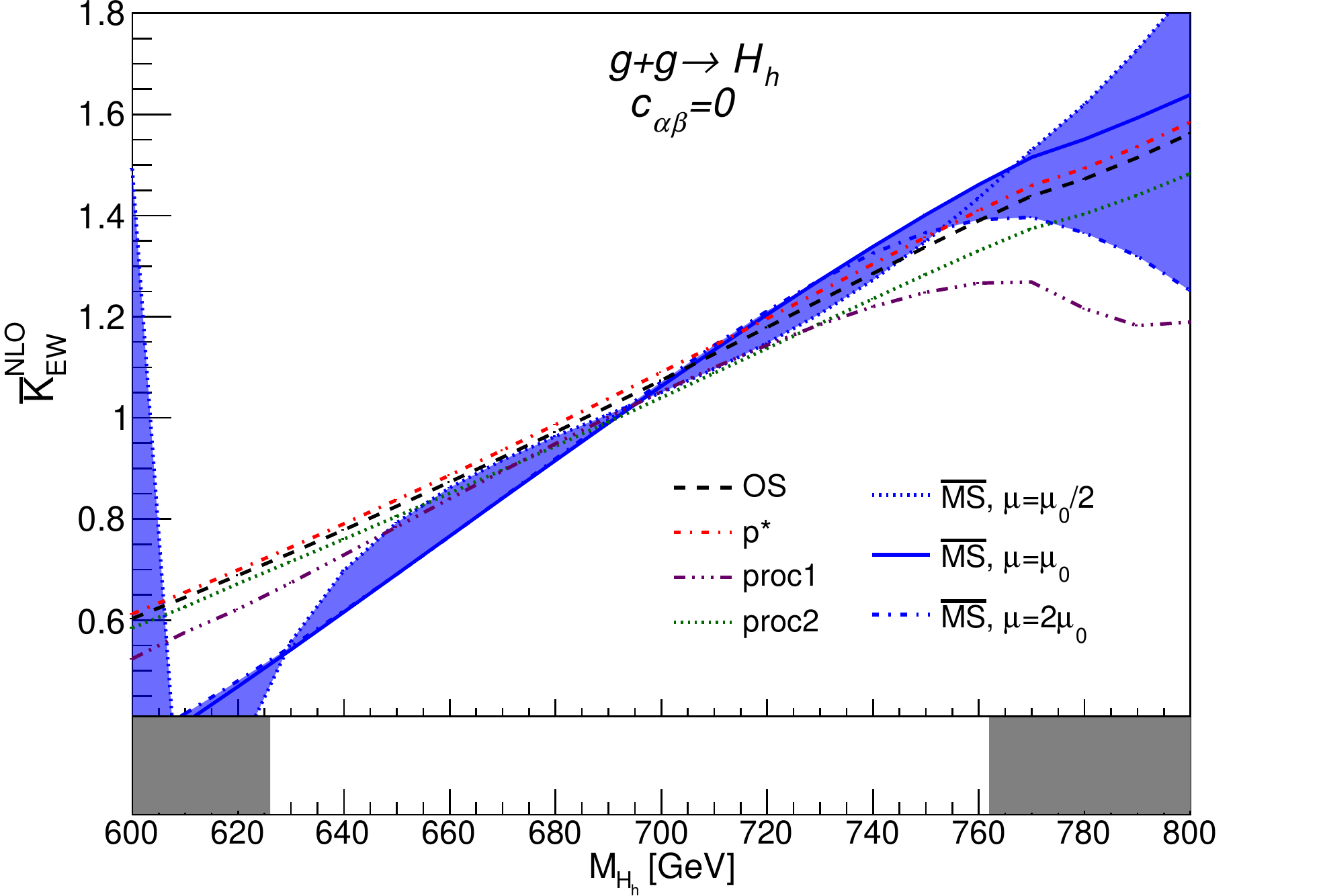}
      \end{center}
    \end{minipage}\\[0.35cm]
    \begin{minipage}{13cm}
      \begin{center}
        \includegraphics[width=12.85cm,bb=0 0 509 379]{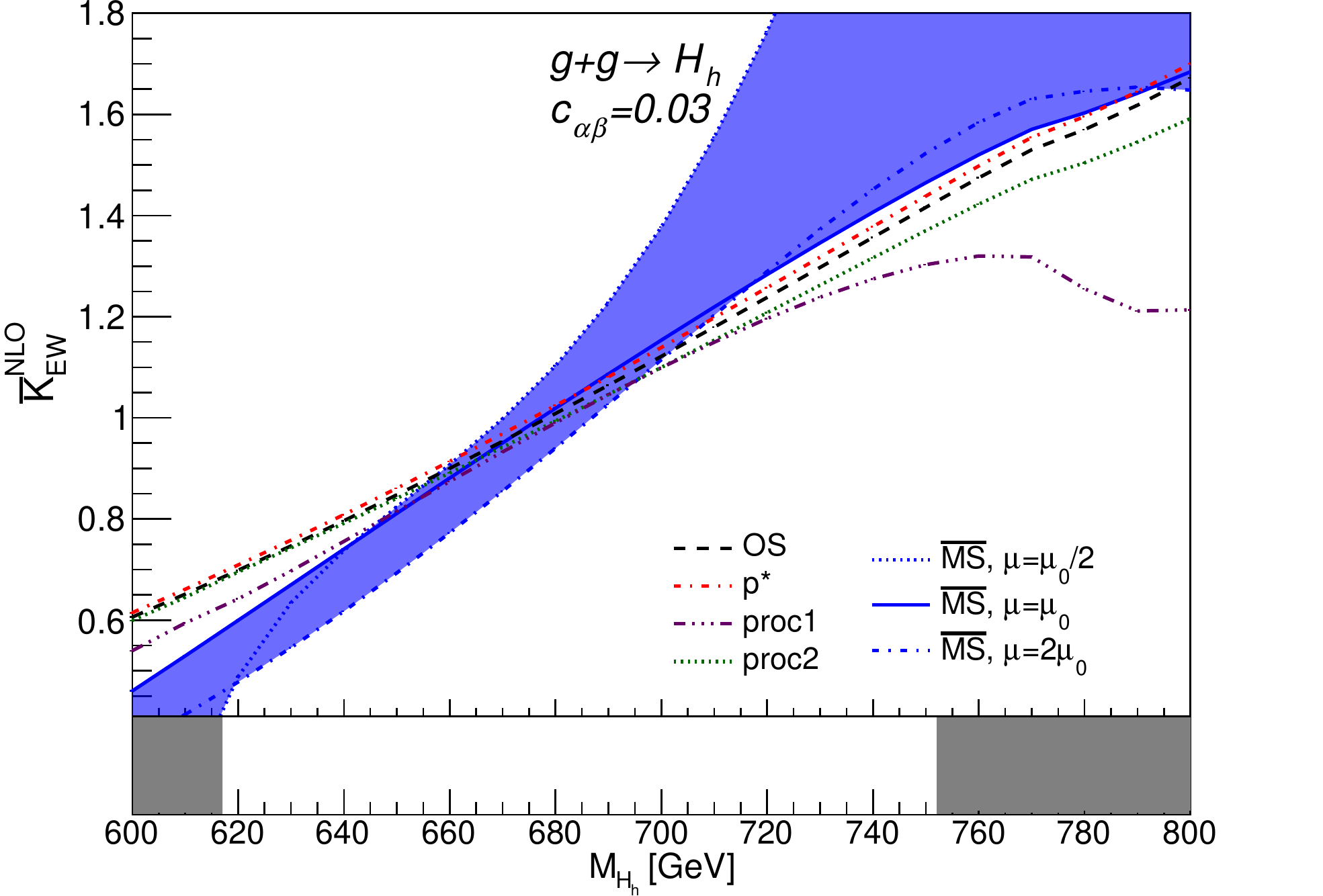}
      \end{center}
    \end{minipage}
  \end{center}
  \caption{Percentage correction of the process $g+g\to\hh$ in various
    schemes. The grey shaded band denotes the region where at
        least one of the couplings $|\lambda_i|/(4\pi)$ becomes larger than
        0.5 and where  one slowly starts to enter in the non-perturbative
        regime.\label{fig:scaleggHh}}
\end{figure}
Fig.~\ref{fig:scaleggHh} shows the K-factor of the NLO
electroweak corrections in different renormalization schemes
for the process $g+g\to\hh$ in the \THDM\
as a function of the heavy, neutral Higgs-boson mass for the same
scenarios as in Fig.~\ref{fig:scaleggHl}.
Again, as for the benchmark points, we use $\kbew$ as the K-factor for
heavy Higgs-boson production.
First of all, we can expect the NLO corrections
for heavy Higgs-boson production to be larger: The LO contribution is
suppressed since for $\tb=2$ and
close to the alignment limit, the coefficient
$\chh\approx-0.5$ is considerably smaller than $\chl\approx1$.
In addition, the LO depends on the heavy Higgs-boson mass $\Mhh$, which
is not fixed in the scenario under consideration. 
In general, we can see that for $|\lambda_i^\mathrm{\tiny max}|/(4\pi)<0.5$,
the K-factors do no longer lie roughly between
$0.99$ and $1.12$ as for light Higgs-boson production, but 
now we observe a larger range of K-factors between $0.4$ and values larger than $1.8$.
Just as for light Higgs-boson production, the proc1 scheme 
differs considerably from the other renormalization schemes
for large $\Mhh$.
The behaviour of the results for the $\MSbar$ renormalization
seems to be more similar to the other schemes when compared with the light
Higgs-boson production, even though there are sizable numerical differences
in some regions. However, this scheme strongly depends on the
definition of the running procedure requiring a more detailed
analysis, which is performed in Appendix~\ref{app:scaledep}.

As expected, due to cancellations among the finite counterterms, 
the \pOS\ and $p^*$ scheme lead to small perturbative corrections. 
The differences between these two schemes may be too small to give an 
estimate of missing higher-order uncertainties. 
As long as $\Mhh$ does not become too large the two process-dependent schemes also perform very well for the 
analyzed scenarios, both for light as well as for heavy Higgs-boson 
production.

\newpage
\section{Summary and conclusion\label{sec:summary}}
We have computed the two-loop electroweak corrections to the production
of a light and a heavy neutral, scalar Higgs boson through gluon fusion
in the \THDM. We have renormalized the new Higgs-boson masses in the
on-shell scheme and provide the electroweak percentage correction in
different renormalization schemes for the mixing angles $\alpha$ and
$\beta$ for these two processes. In particular, for the
mixing angles we have employed the on-shell, $p^*$, $\MSbar$ and two process dependent schemes.  We can
determine the next-to-leading order, electroweak percentage correction
for essentially any scenario of the new mass parameters and the mixing
angles of the CP-conserving  \THDM. In particular, we have computed the two-loop
electroweak corrections for benchmark points collected by the LHC Higgs
cross section working group as well as for individual other example
scenarios.  For Higgs-boson production through gluon fusion, the on-shell scheme performs well for
all chosen scenarios.  
The $\MSbar$ scheme 
can suffer from a large scale dependence and can in general not provide reliable
predictions for all
scenarios.
For the
production of the light Higgs-boson, the electroweak corrections are
always moderate in size, i.e.~they are mostly around 5\% compared to the
LO production cross section for the chosen benchmark points, while for the
production of a heavy Higgs-boson, the electroweak corrections strongly
vary depending on the details of the selected scenario. We have
    solved new technical challenges, which was required to accomplish
    this calculation.  
Our results are also directly applicable to determine the electroweak
percentage corrections for the partial decay widths of the light and
heavy neutral, scalar Higgs-boson decay into two gluons within the \THDM.\\

\vspace{2ex}

\noindent
{\bf{Acknowledgments}}\\
We would like to thank Ansgar Denner and Jean-Nicolas Lang for valuable
discussions and numerical comparisons. The work of L.J. and C.S.  was supported by the
Deutsche Forschungsgemeinschaft~(DFG) under
contract STU~615/1-1. 
The work of S.U. was supported in part by the European Commission through 
the “HiggsTools” Initial Training Network PITN-GA-2012-316704.
The computations were performed with the help of
the High Performance Computing (HPC) cluster of the University of
W{\"u}rzburg, DFG project number 327497565. The Feynman diagrams were drawn with the
program~{\tt{Axodraw}}~\cite{Vermaseren:1994je,*Collins:2016aya}.

\cleardoublepage
\begin{appendix}
\section{Scale dependence of the percentage correction in the $\MSbar$
  scheme without running of {\boldmath{$\cab$}} and {\boldmath{$\tb$}}\label{app:scale}}
The NLO EW percentage
corrections $\delta_{\mbox{\scriptsize EW}}^{\mbox{\scriptsize{NLO}}}$
to the LO partonic cross section are defined through $\hat{\sigma}^{\mbox{\tiny{NLO}}}=
\hat{\sigma}^{\mbox{\tiny{LO}}} (1+\delta_{\mbox{\scriptsize
    EW}}^{\mbox{\scriptsize{NLO}}})$.  
The scale dependence of the percentage correction in the $\MSbar$ scheme
for the process $g+g\to\hl$ in the alignment limit ($\cab=0$) has already
been presented in Ref.~\cite{Denner:2017vms}. It reads
\begin{eqnarray}
\delta^{\mbox{\scriptsize{NLO}},\mu\mathrm{{-}dep.}}_{\mbox{\scriptsize EW}} &=&
{\Gf\*\sqrt{2}\over8\*\pi^2\*\tb^2\*\Mhh^2\*(\Mhh^2-\Mhl^2)}\ln\!\left({\mu^2\over\Mhl^2}\right)
\nonumber\\
&\times&\biggl\{(1-\tb^2)\*(\Mhh^2-\Msb^2)\*\Bigl[3\*\Mhh^2\*\Mhl^2 +
\Msb^2\*(\Mha^2+2\*\Mhc^2-3\Mhh^2)\Bigr] 
\nonumber\\
&+&6\*\mt^2\*(\Mhh^2\*\Mhl^2-4\*\Msb^2\*\mt^2) \biggr\}.
\label{eq:scaledep_gghl}
\end{eqnarray}
The corresponding scale dependence for the process $g+g\to\hh$ in the
alignment limit ($\cab=0$) is given by 
\begin{eqnarray}
\delta^{\mbox{\scriptsize{NLO}},\mu\mathrm{{-}dep.}}_{\mbox{\scriptsize EW}}&=&
{\Gf\*\sqrt{2}\over
  16\*\pi^2\*\tb^2\*\Mhh^2\*(\Mhh^2-\Mhl^2)}\*\ln\!\left(\frac{\mu^2}{\Mhl^2}\right)
\nonumber\\
&\times&
\Biggl\{
  (1-\tb^2)\*(\Msb^2-\Mhh^2)\*\biggl[
    (\Mha^2 + 2\*\Mhc^2)\*\Big[ (1+\tb^2)(\Mhh^2 - \Mhl^2) + 2\tb^2\*\Msb^2 \Big]
\nonumber\\
&&\qquad\qquad\qquad\qquad\qquad
  + 3\*\Mhh^2\*\Big[ \Mhh^2 - \Mhl^2 + \tb^2\*(\Mhh^2 + \Mhl^2 - 2\Msb^2) \Big]
  \biggr]
\nonumber\\
&&\quad
- \,6\*\mt^2\*\Mhh^2\*\bigg[ \Mhh^2 - \Mhl^2 + \tb^2\*( \Mhh^2 + \Mhl^2 ) \bigg]
\nonumber\\
&&\quad
+ 24\*\mt^4\*\bigg[ (1+\tb^2)(\Mhh^2 - \Mhl^2) + \*2\tb^2\*\Msb^2 \bigg]
\Biggr\}.
\label{eq:scaledep_gghh}
\end{eqnarray}
The scale dependence of the process $g+g\to\hh$ in the anti-alignment
limit ($\sab=0$) can be obtained from Eq.~(\ref{eq:scaledep_gghl}) by
interchanging the light and heavy Higgs-boson masses,
i.e. $\Mhl\leftrightarrow\Mhh$. Likewise one can obtain the scale
dependence of the process $g+g\to\hl$ in the anti-alignment limit
($\sab=0$) from Eq.~(\ref{eq:scaledep_gghh}) by again interchanging the light
and heavy Higgs-boson masses. The limit of two equal neutral, scalar
Higgs-boson masses, $\Mhh\equiv\Mhl$, does not exist in all the above
cases due to the denominator structure of Eqs.~(\ref{eq:scaledep_gghl})
and (\ref{eq:scaledep_gghh}).

In order to study and judge the magnitude of the scale dependence,
we compare the size of the coefficient of the scale dependent logarithm,
which we define by
\begin{equation}
\delta^{\mbox{\scriptsize{NLO}},\mu\mathrm{{-}dep.}}_{\mbox{\scriptsize EW}}(x)=
d(x)\ln\!\left(\frac{\mu^2}{\Mhl^2}\right),\;\mbox{with}\quad x=g+g\to H_{l}\quad
\mbox{or}\quad x=g+g\to H_{h}.
\end{equation}
In particular for the BPs in the alignment limit, the coefficient
can easily be obtained by Eqs.~(\ref{eq:scaledep_gghl}) and (\ref{eq:scaledep_gghh}). 
The explicit values of the coefficients are shown in
Tabs.~\ref{tab:BPsscalecab0} and~\ref{tab:BPsscalecabn0}. 
\begin{table}
  \centering
  \begin{tabular}{|c|c|c|c|c|c|c|c|c|}
\hline
BP                & $2_{1A}$ & $2_{1B}$  & $2_{1C}$  & $2_{1D}$ & $3_{A1}$ \\\hline
$d(g+g\to H_{l})$  & -0.071 & -0.072 & -0.004 & -0.004 & -0.058\\\hline
$d(g+g\to H_{h})$  &\m0.215 &\m0.218 &\m0.122 &\m0.155 &\m0.864\\\hline
  \end{tabular}
  \caption{The coefficient in front of the scale dependent logarithm of
    the percentage correction is shown for the benchmark points that
    are in the alignment limit~($\cab=0$).\label{tab:BPsscalecab0}}
\end{table}
%
%
\begin{table}
  \centering
  \begin{tabular}{|c|c|c|c|c|c|c|c|c|}
\hline
BP                & a-1     & b-1      & $2_{2A}$ & $3_{B1}$ & $3_{B2}$ & $4_3$   & $4_4$   & $4_5$  \\\hline
$d(g+g\to H_{l})$  & -0.183 & -0.045 & $4\cdot10^{-5}$ & -0.029 &\m0.003 & -0.015 & 0.003&\m0.072\\\hline
$d(g+g\to H_{h})$  &\m0.562 &\m0.283 & -5.672          &\m15.917 & -1.507 &\m0.353 & 0.201 & -0.331\\\hline
  \end{tabular}
  \caption{The coefficient in front of the scale dependent logarithm of
    the percentage correction is shown for the benchmark points that
    are not in the alignment limit~($\cab\ne0$).\label{tab:BPsscalecabn0}}
\end{table}
\noindent
The size of the coefficients reflects directly the magnitude of the
scale dependence  in the $\MSbar$ scheme without taking into account the running of the
    mixing angles $\cab$ and $\tb$.  Coefficients whose
absolute value is larger than one, like for example for the BPs
$2_{2A}$, $3_{B1}$ and $3_{B2}$ for the process $g+g\to H_{h}$, exhibit a
large scale dependence and give rise to large corrections. Small coefficients, like
for example for the BPs $2_{1C}$, $2_{1D}$, $2_{2A}$ and $3_{B2}$ for
the process $g+g\to H_{l}$, exhibit a very small scale dependence.

\section{Behaviour of the coupling constants {\boldmath{$\lambda_i$}}\label{app:scale_plots}}
The parameters $\lambda_i$ ($i=1,...,5$) of the Higgs potential in
Eq.~(\ref{eq:hpgb}) depend on trigonometric functions
of the mixing angles, i.e. $\tb$ and $\cab$ as well as on the masses of the
Higgs-bosons and the soft-breaking scale. The individual parameters read
\begin{eqnarray}
\label{eq:lambda1}
\lambda_1&=& {g^2\over4\*\Mw^2} \* \left[(\cab-\sab\*\tb)^2\*\Mhh^2+(\sab+\cab\*\tb)^2\*\Mhl^2-\tb^2\*\Msb^2\right],\\
\lambda_2&=&{g^2\over4\*\Mw^2\*\tb^2} \* \left[(\sab+\cab\*\tb)^2\*\Mhh^2+(\cab-\sab\*\tb)^2\*\Mhl^2-\Msb^2\right],\\
\lambda_3&=&{g^2\over4\*\Mw^2\*\tb} \* \left[(\sab+\cab\*\tb)\*(\cab-\sab\*\tb)\*(\Mhh^2-\Mhl^2) + \tb\*(2\*\Mhc^2 - \Msb^2)\right],\\
\lambda_4&=&{g^2\over4\*\Mw^2} \* \left[ \Msb^2 - 2\*\Mhc^2 + \Mha^2 \right],\\
\lambda_5&=&{g^2\over4\*\Mw^2} \* \left[ \Msb^2 - \Mha^2 \right],
\label{eq:lambda5}
\end{eqnarray}
with the coupling $g$ and the $W$-boson mass $\Mw$.

\begin{figure}[!h]
  \begin{center}
    \begin{minipage}{10.5cm}
        \includegraphics[width=10.5cm,bb=0 0 782 432]{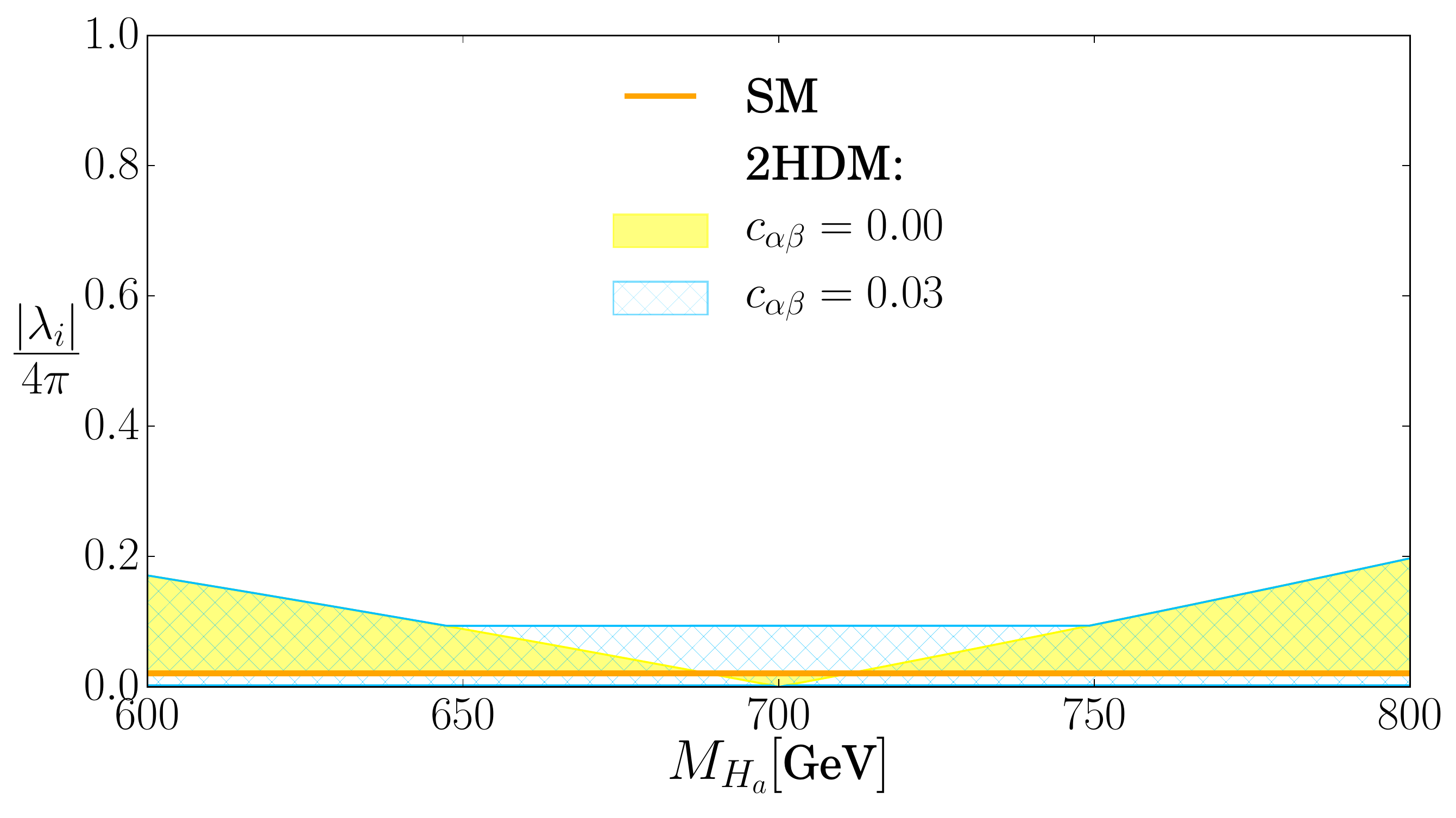}
    \end{minipage}\\[0.5cm]
    \begin{minipage}{10.5cm}
        \includegraphics[width=10.5cm,bb=0 0 782 432]{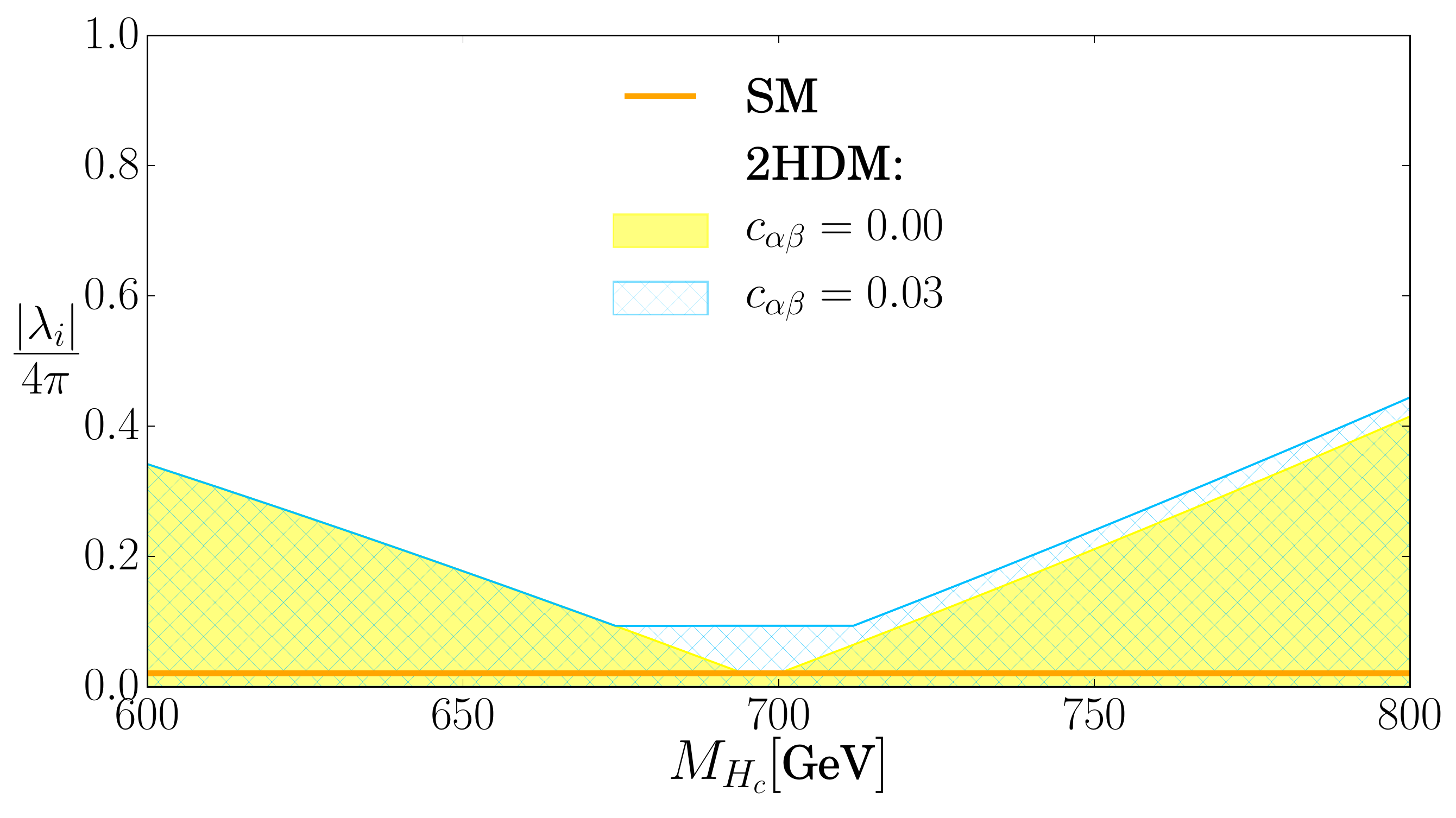}
    \end{minipage}
  \end{center}
  \caption{The range between the minimal and maximal value of the
    modulus of the parameters $\lambda_i/(4\pi)$ as a function of one of
    the new Higgs-boson masses is shown. All other new Higgs-boson mass
    scales are kept fixed at 700~GeV. The range is also compared to the
    SM value $\lambda^{\mbox{\tiny{SM}}}/(4\pi)=0.02$
    (solid, orange line).     The values shown
        here are for the non-$\MSbar$ schemes and thus no running is
        taken into account.
    \label{fig:scan-lambdas-app}}
\end{figure}

Fig.~\ref{fig:scan-lambdas-app} shows the range of
the parameters $|\lambda_i|/(4\pi)$ of the potential for the alternative
scenarios discussed in Section~\ref{sec:results}, where we have fixed
the values of the mixing angles to $\tb=2$ and $\cab=0$ or
$\cab=0.03$. All new heavy mass scales are set to the same value of
700~GeV, except for one Higgs-boson mass, which is varied between 600
and 800~GeV. In the two plots of Fig.~\ref{fig:scan-lambdas-app} the
parameters $|\lambda_i|/(4\pi)$ are shown as a function of $\Mha$ and
$\Mhc$, while the plot as a function of $\Mhh$ is given in
Fig.~\ref{fig:lambda} of Section~\ref{sec:results}. The size of the couplings is
less sensitive to a variation of $\Mha$ and $\Mhc$ than to a variation of $\Mhh$ as shown in
Fig.~\ref{fig:lambda}.

\begin{figure}
  \begin{center}
    \includegraphics[width=14cm,bb=0 0 745 392]{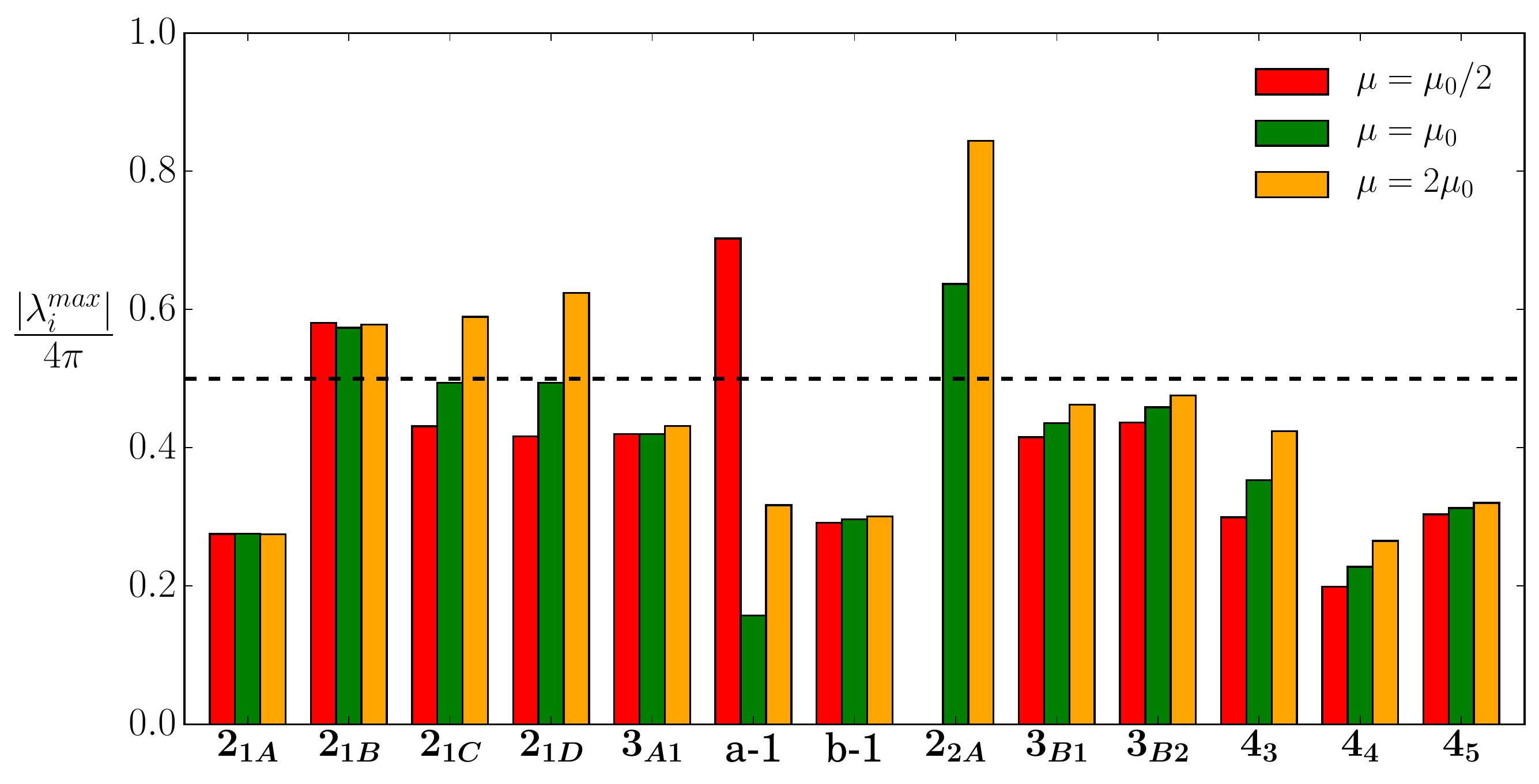}
  \end{center}
  \caption{The running of the largest coupling constants
    $|\lambda_i^\mathrm{\tiny max}|/(4\pi)$  is shown for all
    benchmark points of Section~\ref{sec:results}.  For the
        benchmark point $2_{2A}$ the bar for $\mu=\mu_0/2$ is not given,
        since the RGEs~(\ref{eq:running}) do not allow for a stable solution.
    \label{fig:lambdas_BPs}}
\end{figure}

In Fig.~\ref{fig:lambdas_BPs} we show a bar chart, where we illustrate
the maximal size $|\lambda_i^\mathrm{\tiny max}|/(4\pi)$ of the
couplings of the Higgs-boson potential of Eq.~(\ref{eq:hpgb}) for an
$\MSbar$ renormalization of the mixing angles for the BPs of
Tabs.~\ref{tab:BPcab0} and~\ref{tab:BPcab}. The horizontal, dashed line
shows the location of the value $|\lambda_i^\mathrm{\tiny
  max}|/(4\pi)=0.5$.  The values $|\lambda_i^\mathrm{\tiny max}|/(4\pi)$
are determined with Eqs.~(\ref{eq:lambda1})-(\ref{eq:lambda5}) of
Appendix~\ref{app:scale} and are considered to be defined at the
scale $\mu_d=\mu_0$ of Eq.~(\ref{eq:mu0}) here. The green central bars of
$|\lambda_i^\mathrm{\tiny max}|/(4\pi)$ coincide with the values given
in Tabs.~\ref{tab:BPcab0} and~\ref{tab:BPcab}. In the $\MSbar$ scheme
these central values are then run with the help of the
RGEs~(\ref{eq:running}) to the scale $\mu_0/2$ and $2\mu_0$, and
    they are used to obtain the red and yellow bar, respectively.

\section{$\MSbar$ results in the $\boldsymbol{M^*}$ benchmark scenarios\label{app:scaledep}}
\begin{figure}[!h]
  \begin{center}
    \begin{minipage}{10.6cm}
        \includegraphics[width=10.6cm,bb=0 0 507 379]{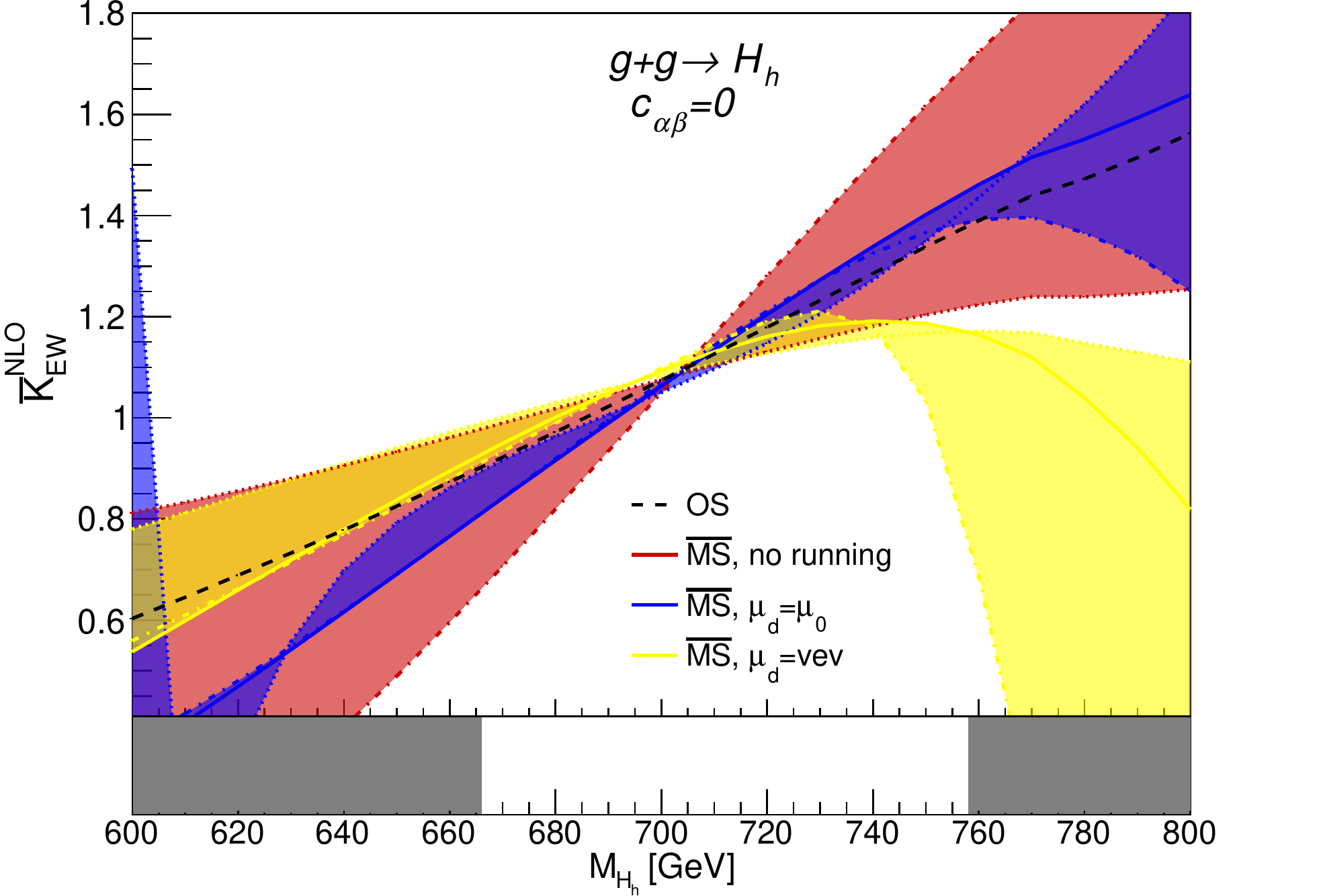}
    \end{minipage}\\[0.55cm]
    \begin{minipage}{10.6cm}
        \includegraphics[width=10.6cm,bb=0 0 507 379]{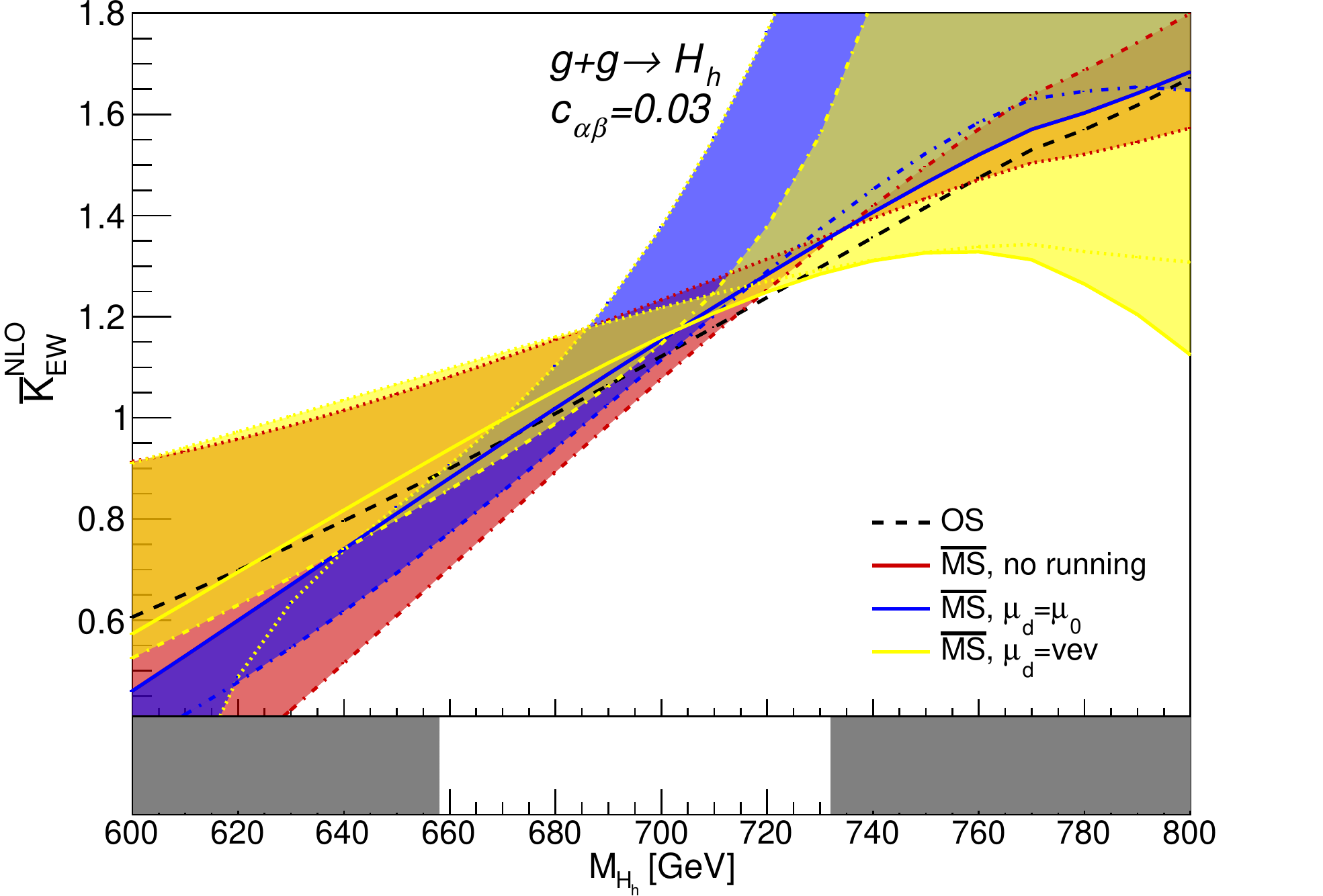}
    \end{minipage}
    \end{center}
\caption{The K-factors of the process $g+g\to\hh$ in various $\MSbar$
  schemes are compared to the on-shell scheme.  Solid lines are for
  $\mu=\mu_0$, while dotted and dash-dotted lines refer to $\mu=\mu_0/2$
  and $\mu=2\mu_0$, respectively.  The grey shaded band denotes the
  region where at least one of the couplings~$|\lambda_i|/(4\pi)$ ($i=1,...5$) of
  the potential in Eq.~(\ref{eq:hpgb}) becomes larger than 0.5 for at
  least one of the scales $\mu$ considered.  The curve with $\mu=\mu_0$ for the
  case of the $\MSbar$ scheme without running coincides with the $\mu=\mu_0$ curve for
  the $\MSbar$ scheme with running, where the parameters are defined at
  $\mu_d=\mu_0$, since in both cases the parameters have been defined at
  $\mu_0$.
\label{fig:ggHhMSbarComp}}
\end{figure}

In this appendix we give the results in the $\MSbar$ renormalization 
scheme of the mixing angles $\alpha$ and $\beta$, for the K-factors of the 
benchmark scenarios at $M^*=700$~GeV described in Section~\ref{sec:Mstar}. 
The mass parameters read
\[
\Mha = \Mhc = \Msb = M^* = 700\,\text{GeV},
\qquad
\Mhh=600,\dots, 800\,\text{GeV},
\]
and the mixing angles are
\vspace{-.3cm}
\begin{equation}
\tb = 2,
  \qquad\qquad
\cab = 0\,\,\mbox{or}\,\,\cab=0.03,
\end{equation}
see also Eq.~(\ref{eq:Msscenario}).
In particular in Figs.~\ref{fig:ggHlMSbarComp} and \ref{fig:ggHhMSbarComp} we 
compare the scale dependence of the EW corrections in three cases: 
\begin{figure}[!h]
  \begin{center}
    \begin{minipage}{10.6cm}
      \includegraphics[width=10.6cm,bb=0 0 507 379]{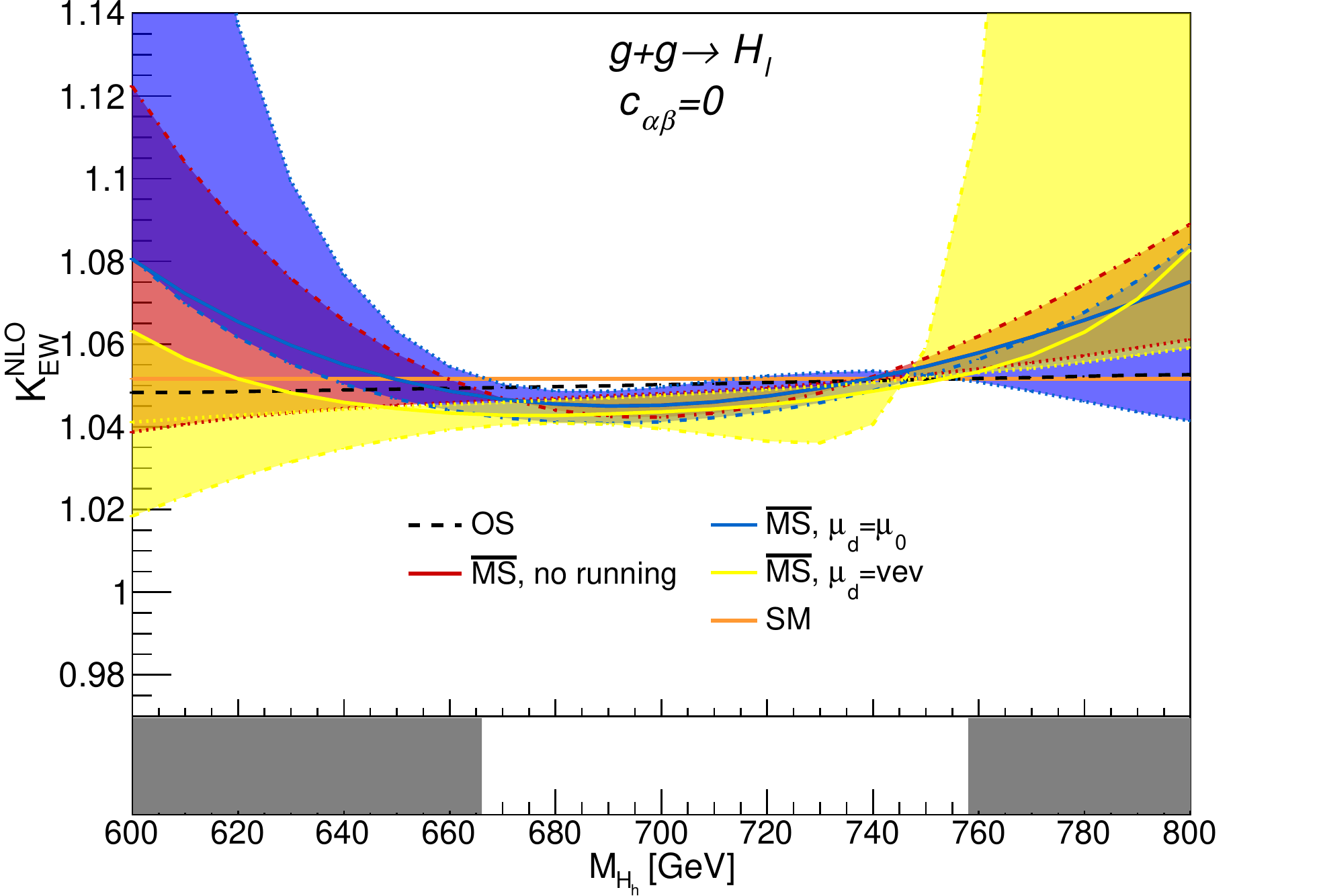}
    \end{minipage}\\[0.55cm]
    \begin{minipage}{10.6cm}
        \includegraphics[width=10.6cm,bb=0 0 507 379]{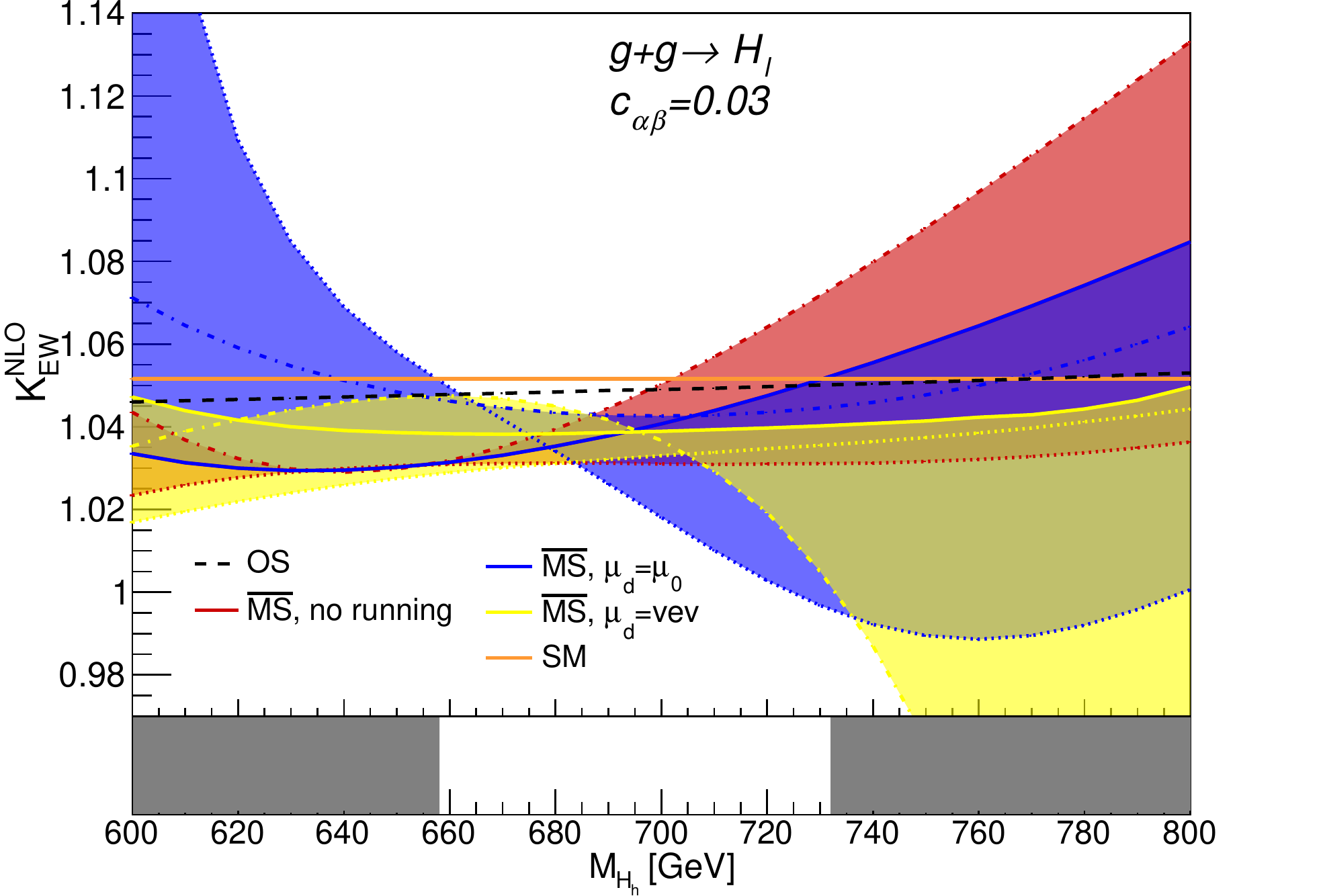}
    \end{minipage}
    \end{center}
\caption{The K-factors of the process $g+g\to\hl$ in various $\MSbar$
  schemes are compared to the on-shell scheme and to the SM result.
  Solid lines are for $\mu=\mu_0$, while dotted and dash-dotted lines
  refer to $\mu=\mu_0/2$ and $\mu=2\mu_0$, respectively.  The grey shaded
  band denotes the region where at least one of the couplings 
  $|\lambda_i|/(4\pi)$ ($i=1,...5$) of the potential in 
  Eq.~(\ref{eq:hpgb}) becomes larger than 0.5 for at least one of the 
  scales $\mu$ considered.  The curve with $\mu=\mu_0$ for the case of the 
  $\MSbar$ scheme without running coincides with the $\mu=\mu_0$ curve for the 
  $\MSbar$ scheme with running, where the parameters are defined at 
  $\mu_d=\mu_0$, since in both cases the parameters have been defined 
  at $\mu_0$.
\label{fig:ggHlMSbarComp}}
\end{figure}
\begin{itemize}
\item The dependence on the renormalization scale $\mu$ enters just in the
loop corrections. The running of the mixing angles
$\alpha$ and $\beta$ is not taken into account (red bands). 
\item The dependence on the renormalization scale $\mu$ enters in the 
loop corrections as well as in the running of the mixing angles $\alpha$
and $\beta$, see Eq.~(\ref{eq:running}); the default scale is
    $\mu_d=\mu_0$, where the soft-breaking scale takes the value
    $\Msb=700$~GeV, and where the mixing angles $\tb$ and $\cab$ take
    the benchmark values $\tb=2$, and $\cab=0$ or $\cab=0.03$ (blue bands).
\item The dependence on the renormalization scale $\mu$ enters in the
loop corrections as well as in the running of the mixing angles $\alpha$
and $\beta$, see Eq.~(\ref{eq:running});  the default scale $\mu_d$ at
which the soft-breaking scale takes the value $\Msb=700$~GeV and at which the
    mixing angles $\tb$ and $\cab$ take the benchmark values $\tb=2$
and $\cab=0$ or $\cab=0.03$ is $\mu_d=$~vev (yellow bands).  
\end{itemize}
The bands in Figs.~\ref{fig:ggHlMSbarComp} and \ref{fig:ggHhMSbarComp} are generated by considering the region of the parametric
space between the three curves corresponding to the values $\mu_0$,
$\mu_0/2$ and $2\mu_0$ for the renormalization scale, where $\mu_0$ is
the typical scale of the processes under consideration as calculated in
Eq.~(\ref{eq:mu0}).  The curve of the
on-shell renormalization scheme (dashed, black curve) is shown as a reference curve
for the scale-independent schemes. For an $\MSbar$ renormalization of the
mixing angles, the couplings $\lambda_i$ of Eq.~(\ref{eq:hpgb}) are scale
dependent. The grey shaded band below each figure shows the region
where at least one of the $\MSbar$ couplings $|\lambda_i|/(4\pi)$ of
Eq.~(\ref{eq:hpgb}) for at least one of the values of $\mu$ becomes
larger than $0.5$ in the considered renormalization scenarios.
Due to the scale dependence, the grey shaded band for an $\MSbar$
renormalization is in general different from the case of considering a
scale-independent scheme for the mixing angles, as shown in
Figs.~\ref{fig:scaleggHl} and \ref{fig:scaleggHh}.  Therefore, the
regions where one enters in the non-perturbative regime can also be
different.

The K-factor for the process $g+g\to\hl$ is shown in
Fig.~\ref{fig:ggHlMSbarComp} for $\cab=0$ (upper plot) and $\cab=0.03$
(lower plot).  In the alignment limit, $\cab=0$, the corrections do not
differ much from the SM in all considered $\MSbar$ schemes when the
couplings $|\lambda_i|/(4\pi)$ remain below $0.5$ (white
region); they are just
    slightly smaller ($1.03<\kew<1.055$).  The
situation changes drastically in the non-perturbative grey region, where
scale variation becomes large and seems to reveal a certain correlation
between perturbativity and scale dependence for this process in the
alignment limit.  This correlation, however, can be a peculiar property of
the chosen scenarios: from Eq.~(\ref{eq:scaledep_gghl}) we see that in
the case of a heavy new scalar sector, the largest contribution to the
scale dependence in the $\MSbar$ scheme comes from the
term proportional to $(\Mhh^2-\Msb^2)$ when the running of the
    mixing angles $\alpha$ and $\beta$ as well as of the soft-breaking
    scale~$\Msb$ is not taken into account; the couplings $\lambda_i$ of
Eqs.~(\ref{eq:lambda1})-(\ref{eq:lambda5}) on the
contrary, show a more
complicated dependence on the masses of the neutral sector. In
the case of exactly equal heavy masses in the alignment limit, they
receive the largest contribution from the $(\Mhh^2-\Msb^2)$ difference
in $\lambda_1$.  Moving slightly away from the alignment scenario (lower
plot with $\cab=0.03$), we notice a larger difference from the SM value
and a wider scale dependence, even in the white (perturbative) region.  In
particular, for $\Mhh$ above $700$~GeV, even remaining in a region where
all $|\lambda_i|/(4\pi)$ are below $0.5$, the $\MSbar$ bands
become quickly wider. Due to this behaviour, $\MSbar$ scale
    dependence can not serve as a good estimator of the theoretical
uncertainty in the light, neutral Higgs-boson production away from the
alignment limit.

In Fig.~\ref{fig:ggHhMSbarComp}, we compare the different $\MSbar$
schemes for the production of a heavy, neutral Higgs-boson in the 
alignment limit (upper plot) and slightly away from it (lower plot).
In this case we do not have a comparison with the SM and we base our
considerations on a comparison with the curve of the on-shell renormalization scheme.
Restricting the analysis of the case $\cab=0$ to the white (perturbative) region
we notice a strong 
dependence on $\Mhh$ for all results. In addition they change a lot for the various $\MSbar$ schemes and for
different values of the scale $\mu$.  The situation is similar but less
dramatic for $\cab=0.03$.  Owing to these large differences the $\MSbar$ scheme can not reliably predict the NLO
    contributions to heavy,
neutral Higgs-boson production. 

\clearpage
\providecommand{\href}[2]{#2}\begingroup\raggedright\endgroup

\end{appendix}
\end{document}